\documentclass[fleqn,11pt]{article}
\usepackage{epsfig}

\newcommand{\BQ}{\begin{equation}}
\newcommand{\EQ}{\end{equation}}
\newcommand{\BQA}{\begin{eqnarray}}
\newcommand{\EQA}{\end{eqnarray}}
\newcommand{\be}{\begin{eqnarray}}

\newcommand{\ee}{\end{eqnarray}}
\newcommand{\x}{{\rm x}}
\newcommand{\xx}{{x_\perp}}
\newcommand{\y}{{y_\perp}}

\newcommand{\z}{z_\perp}
\newcommand{\bb}{b_\perp}
\newcommand{\rr}{r_\perp}
\newcommand{\nn}{\nonumber\\ }
\newcommand{\beq}{\begin{eqnarray}}
\newcommand{\eeq}{\end{eqnarray}}

\def\simge{\mathrel{%
   \rlap{\raise 0.511ex \hbox{$>$}}{\lower 0.511ex \hbox{$\sim$}}}}
\def\simle{\mathrel{
   \rlap{\raise 0.511ex \hbox{$<$}}{\lower 0.511ex \hbox{$\sim$}}}}
\def\bigs{\mathrel{
   \rlap{\raise 0.531ex \hbox{$>$}}{\lower 0.531ex \hbox{$<$}}}}

\def\grad{\nabla}                               
\def\del{\partial}                              

\newcommand{\kk}{k_\perp}
\def\FB{Froissart bound }
\def\FBB{Froissart bound}

\begin{document}

\begin{flushright}
SACLAY-T02/074  \\
NSF-ITP-02-44\\hep-ph/0206241 
\end{flushright}
\vspace{1cm}

\begin{center}
{\LARGE\bf Froissart Bound from Gluon Saturation}

\vspace{0.9cm}

{\large Elena Ferreiro$^{\rm a}$, 
Edmond Iancu$^{\rm b,e}$, Kazunori Itakura$^{\rm c,e}$, 
and  Larry McLerran$^{\rm d}$}\\

\vspace{5mm}

{\it $^{\rm a}$~Departamento de F\'{\i}sica de Part\'{\i}culas,
Universidad de Santiago de Compostela,\\
15706 Santiago de Compostela, Spain}

\vspace{0.1cm}
{\it $^{\rm b}$~Service de Physique Th{\'e}orique, CEA/DSM/SPhT,
Unit{\'e} de recherche associ{\'e}e au CNRS, CEA/Saclay, F-91191 
        Gif-sur-Yvette cedex, France}

\vspace{0.1cm}

{\it $^{\rm c}$~RIKEN BNL Research Center, BNL, Upton NY 11973, USA}

\vspace{0.1cm}

{\it $^{\rm d}$~Nuclear Theory Group, Brookhaven National Laboratory,
        Upton, NY 11973, USA  } 

\vspace{0.1cm}

{\it $^{\rm e}$~Institute for Theoretical Physics, 
                  University of California, 
                  Santa Barbara,\\ CA 93106-4030, USA 
        }

\vspace{0.5cm}

\end{center}

\begin{abstract}

We demonstrate that the dipole-hadron cross-section computed from the
non-linear evolution equation for the Colour Glass Condensate saturates
the Froissart bound in the case of a fixed coupling and for a small
dipole ($Q^2\gg \Lambda_{QCD}^2$).
That is, the cross-section increases as the logarithm squared
of the energy, with a proportionality coefficient involving the
pion mass and the BFKL intercept $(\alpha_s N_c/\pi)4\ln 2$. The pion mass
enters via the non-perturbative initial conditions at low energy.
The BFKL equation emerges as a limit of the non-linear evolution
equation valid in the tail of the hadron wavefunction. We provide a
physical picture for the transverse expansion of the hadron with
increasing energy, and emphasize the importance of the colour correlations
among the saturated gluons in suppressing non-unitary contributions
due to long-range Coulomb tails. We present the first
calculation of the saturation scale including the impact parameter
dependence. We show that the cross-section at high energy exhibits 
geometric scaling with a different scaling variable as compared to the
intermediate energy regime. 

\end{abstract}

\newpage

\section{Introduction }
\setcounter{equation}{0}

One of the striking features of the physics of strong 
interaction is that at high
energies, cross sections are slowly, but monotonously, increasing with $s$
($=$ the total center-of-mass energy squared). For instance,
the data for the $pp$ and $p \bar p$ total cross sections at high energy
can be reasonably well fitted by both $s^\delta$ (with $\delta\approx 0.09$)
and $\ln^2 s$, although the $\ln^2 s$ growth appears to be favored
by recent investigations  \cite{BNetal}.

At a theoretical level, it
was proven many years ago that in the limit $s\to\infty$,
the hadronic total cross sections
must rise no faster than\footnote{The coefficient $C$ in front of
$\ln^2 s$ ($\sigma_{tot}(s) \le C\ln^2 s$) has been left unspecified
in the original analysis by Froissart \cite{F}. The upper bound $C=\pi/m_\pi^2$
has been first derived by Lukaszuk and Martin \cite{LLMartin}, from general
assumptions (unitarity, crossing and analiticity)
on the pion-pion scattering amplitude. This value is consistent
with an early argument by Heisenberg \cite{Heisenberg}, in which
the \FB is actually {\it saturated} : $\sigma_{tot}(s)=(\pi/m_\pi^2)\ln^2 s$.}
 $(\pi/m_\pi^2)\ln^2 s$ \cite{F,Martin,LLMartin}.  This is the
{\it Froissart bound}, which is a consequence of very general principles,
such as unitarity, crossing
and analiticity, but does not rely on any detailed dynamical
information. So, in reality, this bound may very well be not saturated,
although the measured cross-sections seem to do so (at least, 
in so far as they are consistent with a squared log
dependence upon the energy).
In fact, a simple, qualitative mechanism realising this $\ln^2 s$ growth
has been proposed by Heisenberg \cite{Heisenberg} already 
before Froissart bound 
has been rigorously proven (a similar argument is briefly discussed
by Froissart \cite{F}).  However, we are not aware of any field-theoretical
implementation of this, or other, argument.
(See also Ref. \cite{Hebecker} for a recent review and more references,
and Ref. \cite{BNetal} for a recent analysis of the data.)

The goal of this paper is to see how the Froissart bound can be consistent with
modern pictures of high energy strong interactions. To keep the discussion
as simple as possible while still encompassing the interesting physics,
we shall consider the scattering of a {\it colour dipole} off
a hadronic target at very high energy. The ``colour dipole'' 
may be thought of as a quark-antiquark pair
in a colourless state, like a quarkonium,
or a fluctuation of the virtual photon in deep inelastic
scattering. More generally, arbitrary hadronic probes can be considered
as collections of ``colour dipoles'' at least in the approximation
in which the number of colours  $N_c$ is large (so that a gluon excitation
can be effectively replaced by a $q\bar q$ pair). For our approximations
to be justified, we shall assume that the dipole is ``small'', in the sense
that it has a small transverse size, $\rr\ll 1/\Lambda_{QCD}$,
or a large transverse resolution: $Q^2\equiv 1/\rr^2\gg \Lambda_{QCD}^2$.

On the particular example of the dipole-hadron scattering, we shall develop
an argument involving both perturbative and non-perturbative features of QCD,
like non-linear quantum evolution, parton saturation, and confinement,
which will lead us to the conclusion that the total cross-section respects,
and even saturates, the Froissart bound (at least, in the case of a fixed
coupling, that we shall exclusively consider in this paper).
In its essence, our argument may be viewed
as a modern version of Heisenberg's original mechanism.
But our main point is to place this mechanism
within the context of our present theoretical understanding, and clarify
to which extent perturbation theory may play a role in determining
this result.
In brief, we shall demonstrate that this mechanism is naturally
realized by the solutions to non-linear evolution
equations derived within perturbation theory, in Refs. \cite{B,K,W,PI},
but with non-perturbative initial conditions at low energy.
{ In particular, our analysis will confirm, clarify and extend the conclusions
reached in an early study \cite{LR90} based on the GLR equation 
\cite{GLR,MQ},
which is a simplified version of the non-linear evolution equations 
that we shall use in this paper.}

The fact that perturbation theory is the appropriate tool to
describe the high energy behaviour of hadronic cross-sections
is by itself non-trivial, and deserves some comment.
The first, and most certainly true, objection to the use of
weak coupling methods comes from the fact that the
\FB involves the pion mass, and this surely must arise from confinement.
And, indeed, this is how the pion mass enters also our calculations:
via the non-perturbative initial condition that we shall {\it  assume}, and
which specifies the impact parameter dependence of the dipole-hadron
scattering amplitude at low energy. 
But starting with this initial condition, we shall then study 
its evolution with increasing energy within perturbation
theory, and show that the resulting cross-section saturates the \FB 
at high energy.

The adequacy of perturbation theory even for such a limited purpose
--- the study of the quantum evolution of the cross-section with $s$ ---
is still non-trivial \cite{Hebecker}, 
and has been in fact disputed in recent papers \cite{KW02}. 
It is first of all clear that ordinary perturbation theory, in the form
of the BFKL equation \cite{BFKL} --- which resums the dominant radiative 
corrections at high energy, but neglects the non-linear effects associated with
high parton densities ---, fails to describe the asymptotic behaviour
at high energy: The BFKL equation predicts a power-law growth for the
total cross section: $\sigma
\sim  {s}^{\omega\bar\alpha_s}$, with $\omega = 4\ln 2$ and
$\bar\alpha_s=\alpha_s N_c/\pi$, which
clearly violates the Froissart bound. Besides, with increasing
energy, the solution to the BFKL equation ``diffuses'' towards smaller
and smaller transverse momenta, thus making the applicability of
perturbation theory questionable.
 
But the objections in Ref. \cite{KW02}
actually apply to the more recently derived {\it non-linear}
evolution equations \cite{B,K,W,PI}, in which non-linear effects cure
the obvious pathologies of the BFKL equation.
(Alternative
derivations of some of these equations, at least in specific limits,
have been given in Refs. \cite{GLR,MQ,Braun,AM01}.
See also Refs. \cite{Cargese,AMCARGESE} for recent reviews and more references.)
Specifically,
the non-linear effects associated with the high density
of gluons in the hadron light-cone wavefunction (the ``Colour Glass
Condensate'' \cite{PI}) lead to {\it gluon saturation} 
\cite{GLR,MV94,AM2,SAT},
with important consequences for the high-energy scattering: 
First, this introduces
a hard intrinsic momentum scale, the ``saturation scale'',
which is a measure of the density of the saturated gluons in
the impact parameter space, and grows like a power of the energy.
This limits the infrared diffusion \cite{Motyka}  
and thus provides a better justification
for using weak coupling methods at high energy. Second, this ensures
the unitarization of the dipole-hadron scattering amplitude 
at {\it fixed} impact parameter \cite{AM0,AM3,NZ91,BH,K}. 
That is, with increasing energy, 
the hadron eventually turns ``black'' (i.e.,
the dipole is completely absorbed, or the scattering amplitude reaches
the unitarity limit) at any given point $\bb$ in the impact parameter
space.

However, by itself, the unitarization at given $\bb$ is not enough to
guarantee the \FB for the total cross-section, which involves an
integration over all the impact parameters. Indeed, as a quantum mechanical
object, the hadron has not a sharp edge, but rather a diffuse tail,
so, with increasing energy, the ``black disk'' can extend to larger and larger
impact parameters. The radial expansion of the black disk is controlled
by the scattering in the surrounding ``grey area'', where the gluon
density is relatively low and the BFKL equation still applies.
Given the problems of the BFKL equation alluded to before, it is not
a priori clear whether this expansion is slow enough to ensure that
the \FB is respected. This would require a black disk radius which increases
at most logarithmically with $s$. In Heisenberg's original argument (but
using the current terminology), such a
logarithmic increase was ensured by a compensation between the
power-law increase of the scattering amplitude with $s$ and its
exponential decrease with $b\equiv |\bb|$. Such an exponential fall-off
at large impact parameters is, of course, a true
property of {\it full} QCD, and also the crucial assumption about
our {\it initial condition}, but it is not clear  whether this property
can be preserved by the {\it perturbative} quantum
evolution, which involves massless gluons and therefore long range
interactions.

In fact, it was  the main point of Ref. \cite{KW02} to argue
that the Coulomb tails associated with the saturated gluons should replace
the exponential fall-off with $b\equiv |\bb|$ 
of the initial distribution by just a
power-law fall-off, which would be then too slow to ensure the \FBB:
the corresponding black disk radius would increase as a power of $s$.

{ However, as we shall explain in this paper, the argument in Ref. \cite{KW02} 
is irrelevant for the problem at hand, and also incorrect in its original
formulation\footnote{See however the new preprint \cite{KW022} where a modified 
version of this argument has been presented; we shall comment on this new
argument in the Note added at the end of this paper.}. Part of the confusion
in Ref. \cite{KW02} comes from non-recognizing that
the saturated gluons are actually {\it colour neutral} over a relatively
short distance (of the order of the inverse saturation scale) \cite{SAT,Cargese},
and thus cannot produce Coulomb tails at large distances. Rather,
the colour field produced by the saturated gluons
is merely a {\it dipolar} field, whose fall-off
with $b$ is sufficiently fast to respect the Froissart bound for
the scattering of an external dipole.
(In Ref. \cite{KW02}, this was masked by the fact that the authors 
were truly computing the scattering of a {\it coloured} external probe,
although their verbal arguments were formally developed for a ``dipole''.)
As we shall see, for the relevant impact parameters,
this long-range dipole-dipole scattering is
subleading at high energy as compared to the short-range scattering
off the local sources.}

More precisely, we shall demonstrate that the region which controls the
evolution of the cross section is the ``grey area'' outside, but close to
the black disk. { In this area, we expect perturbation theory to apply, since
the local saturation scale is much larger than $\Lambda_{QCD}$.}
For an incoming dipole in this area, the
dominant interactions are those with the non-saturated colour sources
within a ``saturation disk'' (i.e., a disk with radius equal to
the inverse saturation scale) around the impact parameter $\bb$.
This is a consequence of two physical facts: $a\,$) the non-linear 
effects limit the contribution of the distant colour sources, and 
$b\,$) being colourless, the dipole couples only to the local
{\it electric} field (as opposed to the long-range gauge potentials), so it
is less influenced by colour sources which are far away. 

Since, moreover, 
the local saturation length is much shorter than the typical scale 
$\sim 1/\Lambda_{QCD}$  for transverse inhomogeneity in the hadron
(this is where the condition that the dipole is ``small'',
i.e., $Q^2 \gg \Lambda_{QCD}^2$, is essential),
it follows that the quantum evolution proceeds
{\it quasi-locally} in the impact parameter space. { This in turn
implies that, within the grey area, the $\bb$--dependence of the 
scattering amplitude {\it factorizes} out, and is therefore determined
by the initial condition at low energy. On general physical grounds,
we shall assume this initial condition to have 
an exponential fall-off as ${\rm e}^{-2m_\pi b}$ at large 
distances (indeed, pion pairs must control the long distance tail of the 
hadron wavefunction; 
see, e.g., \cite{FK99} and Refs. therein).} Then an argument
similar to the original one by Heisenberg can be used to conclude
that the total cross-section increases like $\ln^2 s$.

The coefficient in front of $\ln^2 s$ in our final result is also interesting,
as it reflects the subtle interplay between the perturbative and 
non-perturbative physics contributing to this result.
Specifically, we shall find that, for {\it any} hadronic target,
\be\label{FB1}
\sigma\,\approx\,\frac{\pi}{2}\,\frac{(\omega \bar\alpha_s)^2}{m_{\pi}^2}
\,\ln^2 s\,\qquad {\rm as}\quad s\to \infty,
\ee
where the pion mass in the denominator enters via the exponential fall-off
of initial condition, while the factor 
$\omega \bar\alpha_s \equiv 4(\ln 2) \alpha_s N_c/\pi$
in the numerator is recognized as the ``BFKL intercept''. This
latter comes up because, in deriving this result, we will have to consider
the solution to the BFKL equation at large energy for fixed $Q^2$.
This is of course the limit for which the BFKL equation has been originally 
proposed \cite{BFKL}, but not also the limit used in more recent
applications of this equation within the context of saturation 
(e.g., in studies of
the saturation scale \cite{AM2,SCALING,MuellerT02}, or of the ``geometric
scaling'' \cite{SCALING})  for a {\it homogeneous} hadron. 

The difference with
Refs.  \cite{AM2,SCALING,MuellerT02} occurs because we consider here
a different physical problem: Rather than studying the quantum evolution
at a {\it fixed} impact parameter --- which would then limit the applicability
of the BFKL equation to not so high energies, such that the
local saturation scale remains
 below $Q^2$ ---, we rather follow the expansion of the
black disk with increasing $s$,
and use the BFKL equation only in the outer grey area at sufficiently
large $b$, where the gluon density remains small even when the energy is large.
In other terms, by increasing the energy
at fixed $Q^2$ {\it and} simultaneously moving 
towards larger impact parameters,
one always finds a corona where the local saturation
scale is much smaller than $Q$, but much larger than $\Lambda_{QCD}$.
For points $\bb$ in this corona, we can prove the factorization of
the scattering amplitude into a $\bb$--dependent ``profile function''
which is determined by the initial condition at low energy, and an
energy-- and $Q^2$--dependent factor which can be computed by solving 
the homogeneous (i.e., no $\bb$--dependence) BFKL equation. 
In the high energy limit at fixed $Q^2$, this calculation
yields the cross-section in eq.~(\ref{FB1}).

In addition to eq.~(\ref{FB1}), we shall find a variety
of new features within our approach.  For example, we shall compute the
impact parameter dependence of the saturation scale, and discover an entirely
nontrivial structure for the radial distribution of matter inside the hadron,
and its evolution with increasing energy.  Related to that, we shall find
that the geometric scaling arguments 
which have been used to characterize deep inelastic scattering at high 
energies become modified by the appearence of two different scales
(associated both with gluon saturation) which
govern the geometric scaling of the total cross-section in different ranges
of the energy. As a result of our analysis, we shall derive an intuitive picture
for the expansion of the hadron in the transverse plane.

The outline of the paper is as follows:  

In the second section, we qualitatively and semi-quantitatively discuss 
how the Froissart bound becomes saturated within the framework of
our knowledge of gluon saturation, non-linear evolution, and its linearized,
BFKL, approximation. This discussion introduces the main arguments to be
demonstrated by the technical developments in the nextcoming sections.

In the third section, we study the properties of a non-linear evolution equation
for the scattering amplitude, the Balitsky-Kovchegov (BK) equation
\cite{B,K}.  We argue that the dominant contribution to the scattering
in the grey area comes from virtual dipoles whose size is much smaller
then the saturation length. This leads us to conclude that the 
corresponding scattering amplitude factorizes in the way alluded
to before. In the rest of the third section, we explore the physical
significance of this factorization by
using the efective theory for the Color Glass Condensate \cite{PI,Cargese}.
We show that an essential ingredient for factorization and Froissart bound
is the colour-neutrality of the saturated gluons within the black disk.

In the fourth section, we compute the radius of the black disk using the 
solution to the BFKL equation.  Then, we extend our results
to compute the impact parameter dependence of the saturation momentum.
We identify two range of values of the impact parameter at which also
the saturation scale factorizes, i.e., it is the product of an
exponentially decreasing function of $\bb$ times a factor increasing
like a power of $s$. These two ranges correspond to the two limits of
the BFKL solution alluded to before: For $\bb$ sufficiently close to the
center of the hadron, the increase with the energy is the same as
for the saturation scale of a {\it homogeneous} hadron, previously 
studied in Refs. \cite{AM2,SCALING,MuellerT02}. This 
comes from a solution to the BFKL equation
in the intermediate regime where $\ln Q^2 \sim \alpha_s\ln s$.
On the other hand, near the edge of the hadron, the increase with $s$
is rather controlled by the high energy  solution at $\ln Q^2 \ll  \alpha_s\ln s$.

In the fifth section, we discuss the implication of the different behaviours 
found at small and large impact parameters for geometrical scaling in
deep inelastic collisions.

In the last section, we summarize our results and present our conclusions.

\section{Saturation scale and the \FB }
\setcounter{equation}{0}

The  dipole-hadron collision will be considered in a special frame,
the ``dipole frame'' \cite{AM0,Cargese}, in which the physical interpretation
of our results becomes most transparent: This is the frame in which the effects
of the quantum evolution are put solely in the wavefunction of the hadron, 
which carries most of the total energy. This being said, it should be stressed
that our final results are independent of this choice of the frame
--- although their interpretation may look different in other frames ---
since at a mathematical level they are based on boost invariant
equations, namely, the non-linear evolution equation  for the scattering
amplitude \cite{B,K,PI} and its linearized, BFKL \cite{BFKL}, approximation.
In fact, the relevant non-linear equation --- namely, the Balitsky-Kovchegov
(BK) equation to be presented in Sect. 3 ---
has been independently derived in the hadron rest frame \cite{B,K},
where the evolution refers to the incoming dipole wave-function, 
and in the infinite momentum frame (which for the present
purposes is equivalent to the ``dipole frame'' alluded to before),
from the evolution of the ``Colour Glass Condensate'' \cite{PI}.

So, let us consider an incoming dipole of transverse size $\rr$, 
with $Q^2\equiv 1/\rr^2\gg \Lambda_{QCD}^2$,
 which scatters off the hadron target
at large invariant energy squared $s$, or rapidity gap $\tau= \ln(s/Q^2)$.
In the dipole frame, most of the total energy is carried
by the hadron, which moves nearly at the speed
of light in the positive $z$  direction. Moreover, any further increase in the 
total energy is achieved by boosting the  hadron alone.
Thus, the dipole rapidity $\tau_{dipole}$ is constant, 
and chosen such as $\alpha_s\tau_{dipole}\ll 1$, so that we can neglect 
higher Fock space components in dipole wavefunction: the dipole is just a
quark-antiquark pair, without additional gluons. On the other hand,
the hadron wavefunction has a large density of small--$\x$ gluons,
which increases rapidly with $\tau$. Here, x is the longitudinal
momentum fraction of the gluons which participate in the collision,
and is related to the rapidity $\tau$ as $\tau=\ln(1/\x)$.

In this special frame, the unitarization effects
in the dipole-hadron
collision can be assimilated to the saturation effects in
the target wavefunction. This is so since 
it is the same momentum scale, namely the {\it saturation scale} 
$Q^2_s$, which sets the border for both types of effects. 

A priori, $Q_s$ is an intrinsic scale of the hadron, 
proportional to the gluon density in the transverse plane
at saturation \cite{GLR,MQ,MV94,Cargese} :
\be\label{Qs}
Q^2_s(\tau,\bb)\,\sim\,\frac{\alpha_s N_c}{N^2_c-1}\,
\x G(\x,Q^2_s,\bb).\ee
In this equation, 
\be
\x G(\x,Q^2,\bb)\,\equiv\,\frac{dN}{d\tau d^2\bb}\ee
is the number of gluons with longitudinal momentum fraction x
and transverse size $\Delta x_\perp \sim 1/Q$ per unit rapidity
and per unit transverse area, at transverse location $\bb$.
The more standard gluon distribution $\x G(\x,Q^2)$ is obtained
by integrating $\x G(\x,Q^2,\bb)$ over all the points  $\bb$
in the transverse plane.
Note that the distribution in $\bb$ is indeed a meaningful quantity
since we consider gluons with relatively large transverse momenta,
$Q^2\gg \Lambda_{QCD}^2$, which are therefore localized over distances
$\Delta x_\perp$ much smaller than the typical scale for
transverse inhomogeneity in the hadron, namely, 
$\Delta\bb \sim 1/\Lambda_{QCD}$.

The saturation scale separates between two physical regimes:
At high transverse momenta $\kk\gg Q_s(\tau,\bb)$, we are in the standard, 
perturbative regime: The gluon density is low, but it increases very fast,
(quasi)exponentially with $\tau$, according to 
linear evolution equations like BFKL \cite{BFKL} or DGLAP \cite{DGLAP}.
At low  momenta $\kk\simle Q_s(\tau,\bb)$, the non-linear effects
are strong even if the coupling is weak,
and lead to {\it saturation}:
The gluon  phase-space density is parametrically large, 
$dN/d\tau d^2\kk d^2\bb\sim 1/\alpha_s$,  but increases
only linearly with $\tau$ \cite{AM2,SAT,Cargese}.

From eq.~(\ref{Qs}), one expects the saturation scale to
increase rapidly with $\tau$, so like the gluon distribution at high momenta.
For a hadron  which is {\it homogeneous} in the transverse plane
 (no dependence upon $\bb$), the $\tau$--dependence of $Q_s$ 
is by now well understood \cite{AM2,SCALING,MuellerT02,LT99,AB01,Motyka},
and will be reviewed in Sect. 4 below. 
Understanding the $\tau$-- and $b$--dependences of the saturation scale
in the general {\it inhomogeneous} case is intimately related to the
problem of the high energy behaviour of the total cross-section,
so this will be a main focus for us in this paper.

To appreciate the relevance of the saturation scale for the
dipole scattering, note that a {\it small} dipole, i.e.,
a dipole with transverse size $\rr\ll 1/Q_s(\tau,\bb)$, where
$\bb$ is the impact parameter, couples to the {\it 
electric} field created by the colour sources in the target.
Thus, its scattering amplitude is proportional
to the correlator of two electric fields, which is the same as the
gluon distribution $\x G(\x,Q^2,\bb)$ with $Q^2\sim 1/r_\perp^2$
\cite{Cargese} :
\be\label{nss}
{\cal N}_\tau(r_\perp,\bb)\,\simeq\,\rr^2\,
\frac{\pi^2\alpha_s C_F}{N_c^2-1}\,\x G(\x,1/\rr^2,\bb)\qquad({\rm single\,\,
scattering}).\ee
This equation, together with the solution $\x G(\x,1/\rr^2,\bb)$
to the BFKL equation, predicts an
exponential increase of the scattering amplitude with $\tau$. If
extrapolated at high energy, this behaviour would violate the
unitarity requirement ${\cal N}_\tau\le 1$. However,
eq.~(\ref{nss}) assumes single scattering, so it is valid only
as long as the gluon density is low enough for the condition 
${\cal N}_\tau(r_\perp,\bb)\ll 1$ to be satisfied.
At high energies, where the gluon density is large,
multiple scattering becomes important, and leads to unitarization.
Assuming that the successive collisions are independent, one
obtains \cite{AM0} :
\be\label{GM-DIP}
{\cal N}_\tau(r_\perp,\bb)\,\simeq\,1-{\rm exp}\bigg\{-\rr^2
\frac{\pi^2\alpha_s}{2N_c}\,\,\x G(\x,1/\rr^2,\bb)\bigg\}\qquad
({\rm multiple\,\,scattering}),\ee
which clearly respects unitarity.

So far, this only demonstrates the role of multiple scattering in
restoring unitarity. The deep connection to saturation follows 
after observing that the condition for multiple scattering to be important
---  that is, that the exponent in eq.~(\ref{GM-DIP}) is of order one ---
is the same as the condition (\ref{Qs})
for gluon saturation in the hadron wavefunction. This is natural since
 the dipole is a direct probe of the gluon distribution
in the hadron, so the non-linear effects in the dipole-hadron scattering 
and in the gluon distribution become important at the same scale.
But this also shows that, in the non-linear regime at 
$Q^2\simle Q_s^2(\tau,\bb)$ one cannot
assume {\it independent} multiple scatterings, as in eq.~(\ref{GM-DIP}):
Rather, the dipole scatters {\it coherently} off the saturated gluons,
with a scattering amplitude which,  unlike eq.~(\ref{GM-DIP}),
cannot be related to the gluon distribution (a 2-point function) alone,
but involves also higher $n$-point functions. This amplitude satisfies
a non-linear evolution equation to be discussed in Sect. 3.

But it is nevertheless true that, as suggested by  eq.~(\ref{GM-DIP}),
the scattering amplitude becomes of order one in the saturation regime at
$Q^2\simle Q_s^2(\tau,\bb)$. (This has been verified via both analytic
\cite{LT99,SAT,SCALING} and numerical \cite{LT99,AB01,Motyka} investigations
of the non-linear evolution equation.) Thus, when increasing
the energy at fixed $Q^2\equiv 1/\rr^2$, 
the unitarity limit ${\cal N}_\tau(r_\perp,\bb) =1$ is eventually reached at
any given impact parameter $\bb$ : for the incoming dipole, the hadron
looks locally ``black''. 

However, the unitarization of the  {\it local scattering amplitude} is not
enough to guarantee the \FB for the {\it   total cross-section}. The latter
is obtained by integrating the scattering amplitude over
all impact parameters:
\be\label{sigma}
\sigma(\tau,\rr)\,=\,2\int d^2\bb\,{\cal N}_\tau(r_\perp,\bb)\,.\ee
It is easy to see that difficulties with the \FB can arise
only because the hadron does not have a sharp edge.
Indeed, if in transverse projection the hadron was a disk
of finite radius $R_0$,
then for sufficiently large energy it would become black at
all the points within that disk, and the total cross-section 
would saturate at the geometrical value $2\pi R_0^2$.

But in reality a hadron is a quantum bound state of the strong
interactions, so its wavefunction has necessarily an exponential tail,
with the scale set by the lowest mass gap in QCD, that is, the pion mass.
Specifically, in the rest frame of the  hadron, the distribution of matter is typically
of the Woods-Saxon type:
\be\label{WS}
\rho(r)\,=\,\frac{\rho_0}{1+\exp\left(\frac{r-R_0}{a}\right)},\ee
where $R_0$ is the typical radial size of the hadron under consideration
(this increases as $A^{1/3}$ for a nucleus with atomic number $A$),
while the thickness $a=1/2m_{\pi}$ is universal (i.e., the same for all hadrons).
This latter involves twice the pion mass because of isospin conservation:
At high energy, one probes the gluons in the hadron wavefunction,
and gluons have zero isospin, so they couple to the external probe 
via the exchange of (at least) two pions. It is therefore $2m_{\pi}$ which
controls the exponential fall off of the scattering amplitude, or of
$\rho(r)$, at large distances:
$\rho(r)\propto {\rm e}^{-2m_{\pi}(r-R_0)}$ for $r-R_0 \simge 1/2m_{\pi}$.

Since, moreover, small-x gluons have large longitudinal wavelengths,
a high energy scattering is sensitive only to the distribution integrated
over $z$, that is, to the transverse {\it profile function} :
\be\label{Sb}
S(b)\,\equiv\,\frac{\int dz\, \rho\Big(\sqrt{\bb^2+z^2}\Big)}
{\int dz \,\rho(z)}\,,\ee
($ b\equiv |\bb|$) which is normalized at the center of the hadron: 
$S(b=0)=1$.
Note that, independent of the detailed form  of $\rho(r)$ in the central
domain at $r<R_0$, the function $S(b)$ decreases exponentially\footnote{But
power law corrections to this exponential decrease
may be numerically important \cite{FK99}.},
$S(b)\simeq {\rm e}^{-2m_{\pi}(b-R_0)}$, for $b-R_0\simge 1/2m_{\pi}$. 

Based on these considerations, we shall assume that the scattering amplitude
for the (relatively low energy) dipole-hadron scattering {in the target
rest frame}  --- which is our initial condition 
for the quantum evolution with $\tau$
--- has the following factorized structure:
\be\label{Nt0}
{\cal N}_{\tau_0}(r_\perp,\bb) \,= \,
{\cal N}_{\tau_0}(r_\perp) S(b),\ee 
with $\tau_0\equiv \tau_{dipole}$ 
and the profile function $S(b)$ introduced above.
At low energy and high $Q^2$, the factorization of the $\bb$--dependence
is natural, since consistent with the DGLAP equation (see, e.g., \cite{GLR}).
For what follows, the crucial feature of the initial amplitude (\ref{Nt0}) is its
exponential fall off $\sim {\rm e}^{-2m_{\pi}(b-R_0)}$ at large distances $b\gg R_0$.

Starting with this initial condition, we increase the energy by boosting
the hadron to higher and higher rapidities. Clearly, the gluon distribution
$\x G(\x,1/\rr^2,\bb)$ at any $\bb$ will increase with $\tau$,
and the BFKL approximation suggests that this increase should be exponential.
Thus, even points $\bb$ which were originally far away in the tail of the
hadron wavefunction ($b- R_0\gg 1/2m_{\pi}$), and did not contribute
to scattering at the initial rapidity $\tau_0$, will eventually give
a significant contribution,  and even become black when the energy
is high enough. That is, with increasing $\tau$,
the {\it black disk} may extend to arbitrarily large impact parameters.

At this point, it is useful to introduce some more terminology.
By the ``black disk'' we mean the locus of the points $\bb$ in the transverse
plane at which the unitarity limit ${\cal N}_\tau(r_\perp,\bb) =1$ 
has been reached in a dipole-hadron collision 
at rapidity $\tau$ and transverse resolution $Q^2\equiv
1/\rr^2$. Equivalently, the condition $Q^2 < Q_s^2(\tau,\bb)$ is satisfied
--- i.e., the gluons with transverse momenta $\sim Q$ are saturated --- at all
the points in the black disk. Given the shape of the initial
matter distribution (\ref{WS}) --- which is isotropic
and decreases from the center of the hadron towards its edge --- it is clear that
the  ``black disk'' is truly a disk, with center at $b=0$ and a radius $R(\tau,Q^2)$
which increases with $\tau$ and decreases with $Q^2$. The black disk radius
is determined by any of the two following conditions
(for more clarity, we shall often rewrite
${\cal N}_\tau(Q^2,\bb)\equiv {\cal N}_\tau(\rr=1/Q,\bb)$ in what follows):
\BQ\label{INT_sat}
{\cal N}_\tau(Q^2,\, \bb)\,=\, \kappa\qquad {\rm for}\qquad b\,=\,R(\tau,Q^2),
\EQ
or
\be\label{condR}
Q_s^2(\tau,\bb)\,=\,Q^2\qquad{\rm for}\qquad b\,=\,R(\tau, Q^2),\ee
which are equivalent since, in turn, the saturation scale is defined by:
\BQ\label{INT_sat1}
{\cal N}_\tau(Q^2,\, \bb)\,=\, \kappa\qquad {\rm for}\qquad Q^2=Q_s^2(\tau,\bb).
\EQ
In these equations, $\kappa$ is a number smaller than one, but not 
{\it much} smaller (e.g., $\kappa=1/2$), whose precise value 
is a matter of convention\footnote{The separation between the saturation
regime at small $Q^2$ and the low-density regime at high $Q^2$ 
being not a sharp one,
 there is some ambiguity in defining the borderline
$Q^2 \equiv Q_s^2(\tau,\bb)$. This is fixed by choosing the number $\kappa$ 
in eq.~(\ref{INT_sat1}).}. For qualitative arguments, and also for
quantitative estimates at the level of the approximations to be developed
below, one can take $\kappa=1$.

We shall also need below the 
``edge of the hadron'' at rapidity $\tau$, by which we mean
the radial distance $R_H(\tau)$ at which the saturation scale becomes of
order $\Lambda_{QCD}$ (this would correspond to the black disk seen
by a large dipole with resolution $Q^2\sim \Lambda_{QCD}^2$, e.g., a pion) :
\be\label{RH}
Q_s^2(\tau,\bb)\,=\,\Lambda_{QCD}^2
\qquad{\rm for}\qquad b\,=\,R_H(\tau).\ee

With these definitions at hand, we now return to the discussion of the
\FBB. At sufficiently  high energy, the total cross-section  
is dominated by the contribution of the black disk (this will be verified
in Sect. 5) : $\sigma\simeq \sigma_{BD}$ with
\be\label{sigmaBD}
\sigma_{BD}(\tau,Q^2)\,\equiv \,2\int d^2\bb\,\Theta(R(\tau, Q^2)-b)\,
{\cal N}_\tau(Q^2,\, \bb)\,\approx \,
2\pi R^2(\tau, Q^2).\ee
Thus, the question about the \FB becomes
a question about the expansion of the black disk with  $\tau$ :
To respect this bound, $R(\tau, Q^2)$ must grow at most
linearly with $\tau$.

One can easily construct a ``na\"{\i}ve'' argument giving such a linear
increase (this is similar in spirit to the old argument by Heisenberg
\cite{Heisenberg}): 
Starting with an initial distribution like (\ref{WS}), assume that,
with increasing $\tau$, the gluon density increases {\it in the same way}
at all the points $\bb$ (outside the black disk),
 so that the $\bb$--dependence of the scattering amplitude factorizes out,
 and is fixed by the initial condition:
\BQ\label{FACT}
{\cal N}_\tau(Q^2,\bb)\,\approx\,S(\bb) {\cal N}_\tau(Q^2).
\EQ
Let us furthermore assume that the function ${\cal N}_\tau(Q^2)$ at large
$\tau$ is given by standard perturbation theory, that is, by the
 solution to the BFKL equation at high energy \cite{BFKL} :
${\cal N}_\tau(Q^2)\propto {\rm e}^{\omega\bar\alpha_s\tau}$, where
 $\omega=4\ln 2$ and  $\bar \alpha_s\equiv N_c\alpha_s/\pi$. Under such
(admittedly crude) assumptions, the scattering amplitude at large $\tau$
and $b\gg R_0$ is given by:
\BQ\label{Nfact1}
{\cal N}_\tau(Q^2,\bb)\,\approx\, \sqrt{\frac{\Lambda^2}{Q^2}}\,\,
{\rm e}^{\omega\bar\alpha_s\tau}\,{\rm e}^{-2m_{\pi}b}\,,\EQ
where we have also included the leading $Q^2$--dependence of the asymptotic
BFKL solution at high energy \cite{BFKL}. ($\Lambda^2$ is some arbitrary reference
scale, of order $\Lambda_{QCD}^2$.) This expression together with the saturation
condition (\ref{INT_sat}) imply:
\BQ\label{BDR}
R(\tau,Q^2)\, \approx \, \frac{1}{2m_\pi}\left( \omega \bar\alpha_s \tau 
-\frac12 \ln \frac{Q^2}{\Lambda^2}\right),
\EQ
and the resulting cross-section saturates the Froissart bound indeed:
\be\label{sigmaFC}
\sigma(\tau,Q^2)\,\approx\,
\frac{\pi}{2m_{\pi}^2}\,\left( \omega \bar\alpha_s \tau
-\frac12 \ln \frac{Q^2}{\Lambda^2}\right)^2\,\sim\,
\frac{\pi}{2}\left(\frac{\omega \bar\alpha_s}{m_{\pi}}\right)^2\tau^2\qquad 
{\rm as\,\,\,\,}\tau
\to\infty.\ee

Our main objective in this paper will be to show that the above, seemingly
na\"{\i}ve, argument is essentially correct, and the results in
eqs.~(\ref{BDR}) and (\ref{sigmaFC}) are truly the predictions of the
non-linear evolution equation for 
${\cal N}_\tau(Q^2,\bb)$ at sufficiently large $\tau$.
This is non-trivial since the na\"{\i}ve argument might go wrong for,
 at least, two reasons:

{\it i})  At very high energies, the non-linear effects become important, and
the use of the BFKL equation becomes questionable. For instance, the unitarization
of the local scattering amplitude ${\cal N}_\tau(Q^2,\bb)$ is precisely the result
of such non-linear effects, which are taken into account by replacing the BFKL
equation with the BK equation.

{\it ii}) Although non-linear, the quantum evolution described
by the BK equation remains perturbative, so it involves massless gluons
and long-range effects which could not only invalidate
the factorization property (\ref{FACT}), but also replace the exponential
fall-off of the initial distribution by just a power-law fall-off 
(an eventuality in which the \FB would be violated).

Nevertheless, as we explain now (and will demonstrate in Sect. 3 below), 
none of these two objections apply to the problem of interest.  Indeed:

{\it i}) However large is $\tau$, there exists an outer corona at 
$R(\tau,Q^2)< b < R_H(\tau)$
where the hadron looks still ``grey'', i.e., where 
${\cal N}_\tau(Q^2,\, \bb)\ll 1$ and the BFKL approximation applies.
It is this ``grey area'' which controls the expansion of the black disk,
and therefore the evolution of the total cross-section at high energy.

{\it ii}) The quantum evolution within the grey area
is quasi-local in $\bb$, because of the non-linear effects which
limit the range of the relevant interactions to 
$\Delta \bb \ll 1/Q_s(\tau,\bb)$.

\begin{figure}
\begin{center}
\includegraphics[width=.85\textwidth] {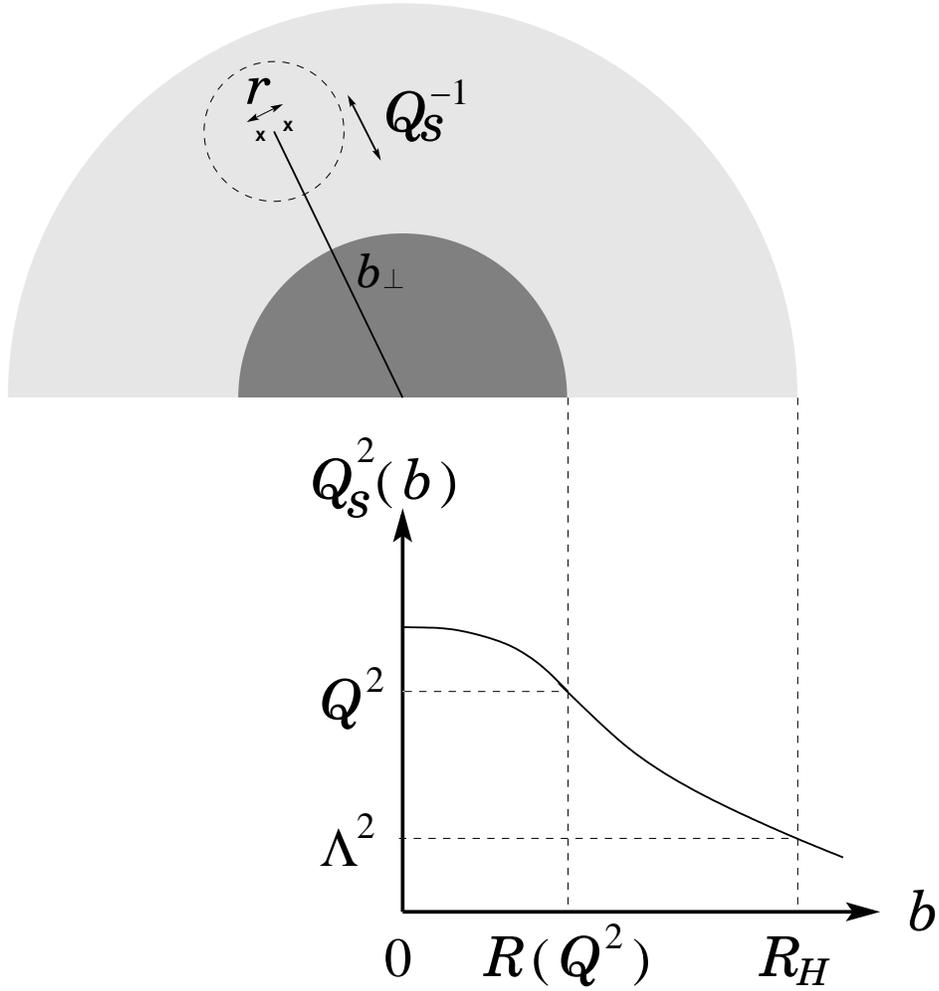}
\caption{A pictorial representation of the
dipole-hadron scattering in transverse projection
 (only half of the hadron disk is shown). The $b$--dependence
of the saturation scale illustrated by the lower plot is the one to be
found in Sect. 4.3.}
\label{BlackD}
\end{center}
\end{figure}

To be more specific, note that, in order to study the expansion
of the black disk with $\tau$, one needs to consider the evolution
of the scattering amplitude ${\cal N}_\tau(Q^2,\bb)$ at points $\bb$
which lie outside the black disk, but relatively close to it. Indeed,
when $\tau\rightarrow \tau+{\rm d}\tau$ with
$\bar\alpha_s {\rm d}\tau \sim 1$ (which is the typical increment
in the high energy regime of interest: $\bar\alpha_s\tau\gg 1$), the
black disk expands by incorporating the points $\bb$ within the range
$R< b < R+{\rm d}R$ with $R\equiv R(\tau,Q^2)$ and ${\rm d}R\sim 1/m_{\pi}$,
cf. eq.~(\ref{BDR}). Such points are sufficiently far away from the black disk
for the local saturation scale to be small compared to $Q^2$ --- that is,
they are in the ``grey area'' ---, but also sufficiently far away from the
edge of the hadron (cf. eq.~(\ref{RH})) for $Q_s(\tau,\bb)$ to be a ``hard''
scale. That is, the following conditions are satisfied for any $\bb$ of interest:
$\Lambda^2_{QCD}\ll Q_s^2(\tau,b)\ll Q^2$. Both inequalities are
important for our argument, as we explain now:

The fact that $Q_s^2(\tau,b)\ll Q^2$ ensures that
the dominant contribution to the evolution
${\rm d}{\cal N}_\tau(Q^2,\bb)$ of the  scattering amplitude 
comes from {\it nearby} colour sources, i.e., from  the sources which
are located
within a saturation disk around $ \bb$ (see also Fig. ~\ref{BlackD}):
\be\label{SR}
|\z-\bb|\,\ll\, \frac{1}{Q_s(\tau,\bb)}\,,\ee
and therefore lie themselves inside the grey area. This is so
because sources which lie
further away are shielded by the non-linear effects.
Besides, being a colour singlet, the dipole is not 
sensitive to the long-range gauge potentials.

The fact that $1/Q_s(\tau,\bb) \ll 1/ \Lambda_{QCD}$ implies that
the transverse inhomogeneity in the hadron can be neglected
when computing the contribution of such nearby sources to
${\rm d}{\cal N}_\tau(Q^2,\bb)$. That is, all the relevant sources
act as being effectively at the same impact parameter, equal to $\bb$.
This explains the factorized expression (\ref{FACT})
for the scattering amplitude. 

Since, moreover, ${\cal N}_\tau(Q^2,\z)\ll 1$ for any
$\z$ satisfying (\ref{SR}), it follows that the function
${\cal N}_\tau(Q^2)$ in eq.~(\ref{FACT})
can be computed by solving the {\it linearized}
(and homogeneous) version of the BK equation, namely the 
BFKL equation without $\bb$ dependence.

Clearly, it was essential for the previous arguments 
that the dipole is ``perturbative'' : $Q^2\gg \Lambda_{QCD}^2$.
This ensures that the separation
$R_H(\tau)-R(\tau,Q^2)$ between the black disk and the hadron edge
(i.e., the width of the ``grey area'')
is sufficiently large for the condition $Q_s^2(\tau,b) \gg\Lambda_{QCD}^2$
to apply at all the points $\bb$ of interest. Besides, one can argue that $Q^2$ is
the scale at which the QCD coupling should be evaluated (see the discussion
in the Conclusions).

A factorization  assumption similar to eq.~(\ref{FACT}) has been
already used in the literature, in particular, in relation with
the \FB \cite{LR90,AGL96}, and also as an Ansatz in the
search for approximate analytical \cite{LT99}
or numerical \cite{LGLM01} solutions to the BK equation.
But in previous work, this assumption  was always based on experience with the
(homogeneous) DGLAP equation, and not duly justified in the small-x
regime. 

That such a factorization is highly non-trivial in the presence of
long-range gauge interactions is also emphasized by a recent
controversy about this point, put forward in Ref. \cite{KW02}.
Specifically, in Ref. \cite{KW02} it has been shown that, as far as the
scattering of a {\it coloured} probe off the hadron is concerned,
the long-range fields created by the saturated gluons provide
a non-unitarizing contribution to the respective cross-section.
On the basis of this example, the authors of Ref. \cite{KW02} have
concluded that the non-linear BK equation provides ``saturation
without  unitarization''. In Sect. 3.2 below, we shall carefully and 
critically examine the arguments in Ref. \cite{KW02},
and demonstrate that, for the physically interesting
case where the external probe is a (colourless) {\it dipole},
there is no problem with unitarity at all. The
{\it long-range} interactions between the incoming dipole and the saturated gluons
give only a small contribution to the scattering amplitude in the grey area,
because the saturated gluons form themselves a dipole (i.e., they are
globally colour neutral), and the dipole-dipole interaction falls off
sufficiently fast with the separation between the two dipoles.
The dominant contribution
comes rather from the {\it short-range} scattering within
the grey area (cf. eq.~(\ref{SR})), for which the factorization
assumption (\ref{FACT}) is indeed justified.

\section{Quantum evolution and black disk radius}\label{QEBD}
\setcounter{equation}{0}

In the effective theory for the Colour Glass \cite{Cargese},
the dipole-hadron scattering is described as scattering of the
$q\bar q$ pair off a stochastic classical colour field which
represents the small-x gluons in the hadron wavefunction.
At high energy, one can use the eikonal approximation to obtain:
\be\label{NS}
{\cal N}_\tau(r_\perp,\bb)\,=\,
1-{\cal S}_\tau(x_{\perp},y_{\perp}),\qquad
{\cal S}_\tau(x_{\perp},y_{\perp})\equiv \frac{1}{N_c}\,
\langle {\rm tr}\big(V^\dagger(x_{\perp}) V(y_{\perp})\big)
\rangle_{\tau},\ee
with $r_\perp=x_{\perp}-y_{\perp}$ the size of the dipole and
$b_\perp=(x_{\perp}+y_{\perp})/2$ the impact parameter
(the quark is at $x_{\perp}$, and the antiquark at $y_{\perp}$).
The $S$-matrix element 
${\cal S}_\tau(x_{\perp},y_{\perp})$ involves the Wilson lines
(path ordered exponentials along the straightline trajectories
of the quark and the antiquark) $V^\dag$ and $V$
built with the colour field of the target hadron. For instance,
\be\label{vtau} 
V^\dagger(x_{\perp})
\,=\,{\rm P}\exp \left \{
ig \int_{0}^{\tau} d{\eta}\,\alpha^a_{{\eta}} (x_{\perp}) t^a\right \},\ee 
where $\alpha_{{\eta}} (x_{\perp})$ is the stochastic 
``Coulomb field'' created by color sources (mostly gluons) at 
rapidities $\tau'<\tau$, and has longitudinal support at (space-time) 
rapidity\footnote{The space-time rapidity is defined as
$\eta\equiv \ln(x^- P^+)$, where $x^-=(t-z)/\sqrt{2}$ is the
light-cone longitudinal coordinate, and $P^+$ is the light-cone
momentum of the hadron. Light-cone vector notations are defined
in the standard way, that is,
$v^\pm\equiv (1/\sqrt 2)(v^0\pm v^3)$. With the present
conventions, the hadron is a right mover, while the dipole
is a left mover.} $\eta\le\tau$. Thus, the integral over $\eta$
in eq.~(\ref{vtau}) is in fact an integral over the longitudinal
extent of the hadron (in units of space-time rapidity) seen by the
external probe in a scattering at rapidity $\tau$. That is,
the actual width of the hadron depends upon the energy of the
collision. This is so since, with increasing energy, gluon modes with larger
and larger longitudinal wavelengths participate in the collision,
so that the hadron looks effectively thicker and thicker \cite{PI}.

The brackets in the definition (\ref{NS})
of the $S$-matrix element refer to the average over all the
configurations of the classical field with some appropriate
probability distribution $W_\tau[\alpha]$ :
\be\label{OBS}
\langle {\rm tr}\big(V^\dagger(x_{\perp}) V(y_{\perp})\big)
\rangle_{\tau}\,=\,
\int\,[{\rm d}\alpha]\,\,{\rm tr}\big(V^\dagger(x_{\perp}) V(y_{\perp})\big)
\,W_\tau[\alpha].\ee
This probability distribution is not known directly, but its variation
corresponding to integrating out gluons in the rapidity window
$(\tau,\tau+d\tau)$ can be computed \cite{JKLW97,PI}.
This leads to a functional evolution equation for $W_\tau[\alpha]$ 
whose precise form  is not needed here (see Ref. \cite{PI} for details). 
Suffices it to say that, via equations like (\ref{OBS}),
the functional equation for $W_\tau[\alpha]$
can be translated into an hierarchy of ordinary evolution
equations for the $n$-point functions of the Wilson lines \cite{PI}.
This procedure yields the same equations as obtained by Balitsky within a different
approach, which focuses directly on the evolution of Wilson line operators
\cite{B}. (The fact that the infinite hierarchy of
coupled equations  by Balitsky can be reformulated as a single
functional  equation has been 
first recognized by Weigert \cite{W}.) In the limit where the number
of colours $N_c$ is large, a closed equation can be written for the
2-point function (\ref{NS})
(with $\bar \alpha_s= N_c\alpha_s/\pi$) :
\beq
	{\partial \over {\partial \tau}} {\cal S}_\tau( x_\perp,\y) & = &
-\bar \alpha_s \int {{d^2z_\perp} \over {2\pi}} 
{{(x_\perp-\y)^2} \over {(x_\perp-z_\perp)^2 (\y-z_\perp)^2}} 
\nonumber \\ &{}& \qquad\qquad\times 
\Big({\cal S}_\tau(x_\perp,\y)- {\cal S}_\tau(x_\perp,z_\perp)
{\cal S}_\tau(z_\perp,\y)   \Big).\label{BK}
\eeq
The same equation has been derived independently by Kovchegov \cite{K} 
within the Mueller's dipole model \cite{AM3}.
We shall refer to eq.~(\ref{BK}) as the Balitsky-Kovchegov (BK) 
equation. 

\subsection{Scattering in the grey area}\label{scat_grey}

In this subsection, we shall study the scattering 
amplitude in the {\it grey} area, and prove the
factorization property (\ref{FACT}).
For more clarity, we shall formulate our arguments
at the level of the BK equation. But one should keep in mind that
our final conclusions are not
specific to the large $N_c$ limit: the same results would have been
obtained starting with the general non-linear evolution equations
in Refs. \cite{B,W,PI}.

Specifically, we shall use eq.~(\ref{BK}) to demonstrate
that the dominant contribution to $\partial {\cal S}_\tau/\partial \tau$
in the grey area comes from {\it short-range} scattering,
i.e. from points $z_\perp$ such that 
\be\label{largez}
|\z-\bb|\,\ll\, \frac{1}{Q_s(\tau,\bb)}\,\ll\,\frac{1}{\Lambda_{QCD}}
\,.\ee

As explained in Sect. 2, the impact parameters of interest are such that
the following inequalities are satisfied (with $Q^2\equiv 1/\rr^2$):
$Q^2 \gg Q_s^2(\tau,b)\gg \Lambda^2_{QCD}$. That is,
the dipole is small not only as compared to the typical
scale for non-perturbative physics and transverse inhomogeneity in the
hadron, namely $1/{\Lambda_{QCD}}$, but also as compared to the
shorter scale $1/Q_s(\tau,\bb)$, which is the local saturation length.
This implies that the dipole is only
weakly interacting with the hadron: ${\cal S}_\tau( x_\perp,\y)\simeq 1$, 
or ${\cal N}_\tau(r_\perp,\bb)\ll 1$. But this does not mean
that we are a priori allowed to linearize  eq.~(\ref{BK}) with respect
to ${\cal N}_\tau\,$. Indeed, the r.h.s. of
this equation involves an integral over all $\z$, so the virtual
dipoles with transverse
coordinates $(\xx,\z)$ or $(\z,\y)$ can be arbitrarily large.
In fact, we shall see below that the dominant contribution comes nevertheless
from $\z$ which is
relatively close to $\bb$, in the sense of eq.~(\ref{largez}),
but the upper limit $1/Q_s(\tau,\bb)$ 
in this equation is a consequence of the non-linear effects.

To see this, it is convenient to divide the integral over $\z$ in eq.~(\ref{BK}) 
into two domains (``short-range'' and ``long-range''):
\BQA
{\rm (A)} \,\,\,\, |\z-\bb|\,\ll\, 1/Q_s(\tau,\bb)\, ,\qquad\ 
{\rm (B)} \,\,\,\, 1/Q_s(\tau,\bb)\,\ll \,|z_\perp-\bb|\, . 
\EQA
It is straightforward to compute the contribution of domain
(B) to the r.h.s. of eq.~(\ref{BK}): In this range, $|\xx-\z|\sim
 |\z-\y|\sim |z_\perp-\bb|$, so the virtual
dipoles are both relatively large, and therefore strongly absorbed. 
Thus, to estimate their contribution,
one can set ${\cal S}_\tau(x_\perp,\z)\approx 0$
and ${\cal S}_\tau(\z,\y)\approx 0$, and approximate the 
(``dipole'' \cite{AM0,AM3,AMCARGESE}) kernel in the BK equation as:
\be\label{dipK}
{{(x_\perp-\y)^2} \over {(x_\perp-z_\perp)^2 (\y-z_\perp)^2}}\,\approx\,
{\rr^2\over (\z-\bb)^4}\,.\ee
This gives (with $u_\perp^2\equiv (\z-\bb)^2>1/Q_s^2(\tau,\bb)$) :
\be\label{IIcon0}
\left.\frac{\del }{\del \tau} {\cal S}_\tau(x_\perp,\y)\right\vert_{\rm (B)}
& \simeq & -\bar \alpha_s \rr^2 {\cal S}_\tau(x_\perp, \y)\int_{1/Q_s^2} 
\frac{d u_\perp^2}{u_\perp^4}\nn
&=& -\frac{\bar\alpha_s}{2}\,\Big(\rr^2 Q_s^2(\tau,\bb)\Big)
{\cal S}_\tau(x_\perp, \y).
\ee
Since ${\cal S}_\tau(x_\perp, \y)$ is of order one for the small
dipole of interest, we deduce the following order-of-magnitude
estimate (which we write for $\del {\cal N}_\tau/\del \tau$, for
further convenience) :
\be\label{IIcon}
\left.\frac{\del }{\del \tau}{\cal N}_\tau(r_\perp, \bb)\right\vert_{\rm (B)}
& \sim & \bar\alpha_s \,\rr^2 Q_s^2(\tau,\bb).\ee
Note that, even for this ``long range'' contribution, 
the integral in eq.~(\ref{IIcon0}) is
dominated by points $u_\perp$ which are relatively
close to the lower limit $1/Q_s(\tau,\bb)\,$; 
this is so because the dipole kernel (\ref{dipK})
is rapidly decreasing at large distances $|\z-\bb|\gg \rr$.

To evaluate the corresponding contribution of domain (A), 
we note first that, in this domain, all the dipoles are small,
so the scattering amplitude is small,
${\cal N}_\tau \ll 1$, for any of them. It is therefore appropriate
to linearize the r.h.s. of eq.~(\ref{BK}) with respect to ${\cal N}_\tau$
(below, $u_\perp=\xx-\z$):
\be\label{BFKL}
\left.\frac{\del }{\del \tau} 
{\cal N}_\tau(r_\perp, \bb)\right\vert_{\rm (A)}
&\! \! \simeq \! \! &
- \bar\alpha_s \int^{1/Q_s} \frac{d^2u_\perp}{2\pi}\,
\frac{\rr^2}{u_\perp^2(\rr-u_\perp)^2}\,\\
&&\! \!\! \!\times \left\{{\cal N}_\tau(\rr,\bb)-
{\cal N}_\tau\Big(u_\perp,\bb-\frac{u_\perp-\rr}{2}\Big)
- {\cal N}_\tau\Big(\rr-u_\perp,\bb-\frac{u_\perp}{2}\Big)\right\}.\nonumber
\label{Kovchegov2}
\EQA
This is recognized as the BFKL equation in the coordinate representation. 
Since both $u_\perp$ and $\rr$ are small as compared to $1/Q_s(\tau,\bb)$,
and therefore much smaller than $1/{\Lambda_{QCD}}$, it is appropriate to
neglect the hadron inhomogeneity when evaluating the r.h.s. of this equation.
That is, all the functions
${\cal N}_\tau$ in the r.h.s can be evaluated
at the same impact parameter, namely $\bb$. 

To obtain an order-of-magnitude
estimate for the r.h.s. of eq.~(\ref{BFKL}), we need an estimate for
the function ${\cal N}_\tau(\rr,\bb)$
in the regime where $\rr\ll 1/Q_s(\tau,\bb)$. An approximate solution
valid in this regime will be constructed in Sect.~\ref{BDFB}.
But for the present purposes, we do not need all the details of this
solution. Rather, it is enough to use the following 
``scaling approximation'' 
\be\label{Nscaling}
{\cal N}_\tau(\rr,\bb)&\simeq& \left(\rr^2Q_s^2(\tau,\bb)\right)^{\lambda},\ee
with $\lambda\le 1$. In Ref. \cite{SCALING}, this approximation has
been justified for a homogeneous hadron (no dependence upon
$\bb$). In Sect. 5 below, we shall find that, in the regime of interest,
geometric scaling remains true also in the presence of inhomogeneity.

The highest value $\lambda=1$ 
corresponds to the ``double logarithmic regime''\footnote{Strictly
speaking, there is no geometric scaling in this regime, but 
for power counting purposes
one can assume that ${\cal N}_\tau(\rr,\bb)\propto
\rr^2 \,\x G(\x,1/\rr^2,\bb)$ (cf. eq.~(\ref{GM-DIP})) is 
linear in $\rr^2$. Indeed, at very high $Q^2$, the gluon
distribution $\x G(\x,Q^2,\bb)$ is only weakly
dependent upon $Q^2$.} in which the
dipole is extremely small, $\ln (Q^2/Q_s^2(\tau,\bb))\gg 1$, 
or, equivalently,
its impact parameter $\bb$ is far outside the black disk, 
$b\gg R(\tau,Q^2)$. Here, however, we are mostly interested in points
$\bb$ which are not so far away from the black disk, since we would like
to study how the latter expands by incorporating points from the grey area.
In this regime, i.e., for $b > R(\tau,Q^2)$ but such that 
$(b-R(\tau,Q^2))/R(\tau,Q^2) \ll 1$, the scattering amplitude
is given by eq.~(\ref{Nscaling}) with a power $\lambda$ which is
strictly smaller than one (see Sect. 5).

To simplify the evaluation of eq.~(\ref{BFKL}), we shall divide domain (A) in two
subdomains, in which further approximations are possible:
(A.I) one of the two virtual dipoles, say  $u_\perp$, is much smaller
than the other one\footnote{That is, in the notations of eq.~(\ref{BK}),
the point $\z$ is within the area occupied by the original dipole
$(\xx,\y)$, and much closer to $\xx$ than to $\y$.}
: $u_\perp\ll \rr$; (A.II) both virtual dipoles
are larger than the original one, although still smaller than one
saturation length: $\rr \ll u_\perp \sim |u_\perp-\rr| \ll 1/Q_s(\tau,\bb)$.

In domain (A.I), the first and third ``dipoles'' in  
the r.h.s. of eq.~(\ref{BFKL}) cancel each other (since
$\rr^2 \approx (\rr-u_\perp)^2$), and we are left with
\be\label{BFKLA}
\left.\frac{\del }{\del \tau} 
{\cal N}_\tau(r_\perp, \bb)\right\vert_{\rm (A.I)}
\, \simeq \,
  2 \bar\alpha_s \int^{\rr} \frac{d^2u_\perp}{2\pi}\,
\frac{1}{u_\perp^2}\,\left(u_\perp^2Q_s^2(\tau,\bb)\right)^{\lambda}
\,\sim\,\bar\alpha_s \left(\rr^2Q_s^2(\tau,\bb)\right)^{\lambda},\ee
where the factor of 2 takes into account that one can choose any of
the two virtual dipoles as the small one.

In domain (A.II), one can neglect ${\cal N}_\tau(\rr,\bb)\ll
{\cal N}_\tau(u_\perp,\bb)\simeq {\cal N}_\tau(u_\perp-\rr,\bb)$,
and obtain:
\BQA\label{BFKLB}
\left.\frac{\del }{\del \tau} 
{\cal N}_\tau(r_\perp, \bb)\right\vert_{\rm (A.II)}
&\simeq & 2 \bar\alpha_s \int^{1/Q_s}_{\rr} \frac{d^2u_\perp}{2\pi}\,
\frac{\rr^2}{u_\perp^4}\,\left(u_\perp^2Q_s^2(\tau,\bb)\right)^{\lambda}
\nonumber\\
&= & \bar\alpha_s \left(\rr^2Q_s^2(\tau,\bb)\right)^{\lambda}\,
\frac{1-(\rr^2Q_s^2(\tau,\bb))^{1-\lambda}}{1-\lambda}\,,
\EQA
which is of the same order as the (A.I)--contribution (\ref{BFKLA})
when $\lambda < 1$, but is logarithmically enhanced over it,
and also over the long-range contribution (\ref{IIcon}), when
$\lambda\to 1$ :
\BQA\label{BFKLB2}
\left.\frac{\del }{\del \tau} 
{\cal N}_\tau(r_\perp, \bb)\right\vert_{\rm (A.II)}\,\sim\,
\bar\alpha_s \left(\rr^2Q_s^2(\tau,\bb)\right)
\ln \frac{1}{\rr^2 Q_s^2(\tau,\bb)}\,\qquad {\rm when}\quad
\lambda=1.
\EQA

By comparing eqs.~(\ref{IIcon}) and (\ref{BFKLA})--(\ref{BFKLB2}), it
should be clear by now that, for any  $\lambda\le 1$, the short-range
contribution, domain (A), dominates over the long-range one, domain (B).
In other words, from the analysis of the non-linear BK equation, we have
found that,
for a ``small'' incoming dipole, the dominant contribution to the quantum 
evolution of ${\cal N}_\tau(r_\perp, \bb)$ 
comes from still ``small'' virtual dipoles 
$\{(x_\perp,\z),(\z,\y)\}$ (see Figure \ref{BlackD}).
 This has two important consequences. 
({\it i}) One can linearize the BK equation with respect 
to ${\cal N}_\tau$, as we did already in eq.~(\ref{BFKL}).
 This gives the BFKL equation. 
({\it ii}) One can ignore the transverse
inhomogeneity in the BFKL equation. That is, one 
can replace eq.~(\ref{BFKL}) by
\BQA\label{BFKLb}
\frac{\del}{\del\tau}{\cal N}_\tau(\rr,\bb)&\simeq& \bar\alpha_s 
\int \frac{d^2\z}{2\pi}\frac{\rr^2}{(\z-x_\perp)^2(\z-\y)^2}\nonumber\\
&&\quad\times \Big\{
{\cal N}_\tau(\z-x_\perp,\bb)+{\cal N}_\tau(\z-\y,\bb)-{\cal N}_\tau(\rr,\bb)
\Big\},
\EQA
in which all the amplitudes ${\cal N}_\tau$ are evaluated at 
the {\it same} impact parameter, namely at $\bb$.

Note that, as compared to  eq.~(\ref{BFKLB}),
there is no need to insert an upper cutoff $\sim 1/Q_s$
in the integral in eq.~(\ref{BFKLb}). This is so since, to the
accuracy of interest, the solution ${\cal N}_\tau(\rr,\bb)$ to
eq.~(\ref{BFKLb}) is actually insensitive to such a cutoff. 
One can understand this on the basis of eq.~(\ref{BFKLB}) : 
For large $\rr$ (with $\rr^2 Q_s^2(\tau,\bb)\ll 1$ though),
the solution has the ``scaling'' behaviour in 
eq.~(\ref{Nscaling}) with a power $\lambda$ which is
strictly smaller than one (see  the discussion in Sects. 4 and 5).
Then the integral in eq.~(\ref{BFKLB}) is dominated by points
$u_\perp$ which are close to the lower limit $\rr$, i.e., by virtual 
dipoles which are not much larger than the incoming dipole. 
The dependence upon the upper cutoff
$\sim 1/Q_s$ is therefore a subleading effect, which can be 
safely ignored.

We finally come to the last step in our argument:
Since the dependence of  eq.~(\ref{BFKLb}) upon $\bb$ is only ``parametric'',
it is clear that the impact parameter dependence of the solution
is entirely fixed by the initial condition. This, together with eq.~(\ref{Nt0}),
 implies that ${\cal N}_\tau(\rr,\bb)$ has the factorized structure:
\BQ\label{Nfact}
{\cal N}_\tau(\rr,\bb)=S(\bb) {\cal N}_\tau(\rr)
\EQ
where $S(\bb)$ is the transverse profile of the initial condition, 
while ${\cal N}_\tau(\rr)$ satisfies the homogeneous BFKL equation 
and will be discussed in Sect.~\ref{BDFB}. A brief inspection of the
previous arguments reveals that the terms neglected in our 
approximations are suppressed by either powers of $\rr^2 Q_s^2(\tau,\bb)$
(e.g., the long-range contribution (\ref{IIcon}), or the cutoff--dependent
term in eq.~(\ref{BFKLB})), or by powers of ${\Lambda^2_{QCD}}/Q_s^2(\tau,\bb)$
(the inhomogeneous effects in eq.~(\ref{BFKL})).
This specifies the accuracy of the factorized approximation in
eq.~(\ref{Nfact}).

\subsection{More on the saturated gluons}\label{more_saturated}

In the previous subsection, we have seen that colour sources located
far away from the impact parameter of the dipole, such as
$|\z-\bb| > 1/Q_s(\tau,\bb) $, do not significantly
contribute to the scattering amplitude in the grey area.
In what follows, we shall examine more carefully a particular
contribution of this type, namely, that associated with 
the saturated gluons within the black disk: $|\z|< R(\tau, Q^2)$.
Indeed, it has been recently argued \cite{KW02} that,
by itself, this contribution would lead to unitarity violations. 
To clarify this point, we shall compute this  contribution
within the effective theory for the Colour Glass Condensate, where
the physical interpretation of the result is transparent.
The same result will be then reobtained  from the
BK equation. Our analysis will confirm that, for
impact parameters $\bb$ within the grey area, this long-range
contribution is indeed subleading, and can be safely neglected  
at high energies. As we shall see, non-unitary contributions of the type 
discussed in Ref.~\cite{KW02} appear only in the physically uninteresting
case where the exernal probe carries a 
non-zero colour charge (as opposed to the colourless dipole).

According to eqs.~(\ref{NS}) and (\ref{vtau}), the  scattering 
amplitude at rapidity $\tau$ depends upon the Coulomb field 
$\alpha_{\eta}^{a}(x_\perp)$ at all the space-time rapidities
$\eta\le \tau$. In general, this is related to the colour sources
in the hadron via the two-dimensional Poisson equation
$-\nabla^2_\perp \alpha_{\eta}^{a}(x_\perp)=\rho_{\eta}^{a}(x_\perp)$,
with the solution:
\be\label{alpha1}
\alpha_{\eta}^{a}(x_\perp)&=&
\int d^2z_\perp 
\langle x_\perp|\frac{1}{-\grad^2_\perp}|z_\perp\rangle\,
\rho_{\eta}^{a}(z_\perp).
\ee
In this equation, $\rho_{\eta}^{a}(x_\perp)$ is the colour charge density
(per unit transverse area per unit space-time rapidity)
of the colour sources at space-time rapidity $\eta$. Here, we are only
interested in such colour sources which are {\it saturated}.
To isolate their contribution, it is important to remark that these sources
have been generated by the quantum evolution up to a ``time''
equal to $\eta$, so the corresponding saturation scale is $Q_s(\eta,\bb)$
(and not $Q_s(\tau,\bb)$). Thus, the integration
in eq.~(\ref{alpha1}) must be restricted to $|z_\perp|\le R(\eta, Q^2)$,
with $R(\eta, Q^2)$ the black disk radius at  rapidity $\eta$ 
(cf.  eq.~(\ref{condR})), and $ Q^2$ the typical momentum carried
by the Fourier modes of $\alpha_{\eta}^{a}(x_\perp)$ (as usual,
this is fixed by the transverse size of the incoming dipole).

\begin{figure}
\begin{center}
\includegraphics[height=11.cm] {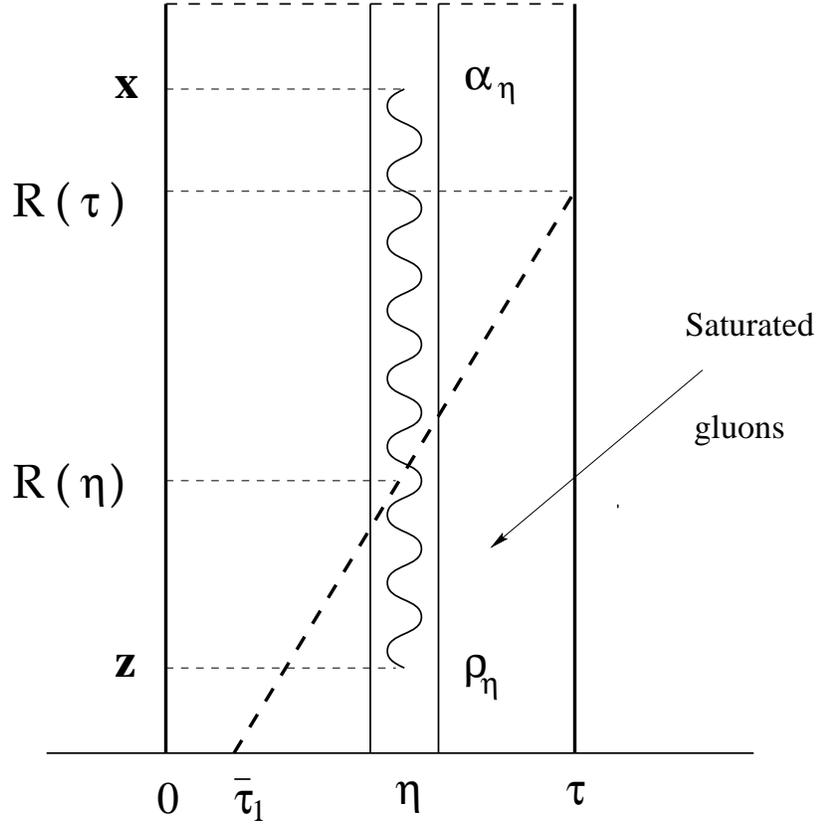}
\caption{The longitudinal profile of the hadron as it appears in
a scattering at given $\tau$ and $Q^2$.
The longitudinal coordinate is on the horizontal axis, and
is measured in units of space-time rapidity. 
The tranverse coordinate is on the vertical axis. A longitudinal
layer at rapidity $\eta$ is delimited for more clarity.
The wavy line represents the colour field $\alpha_\eta$
created at point $\xx$ by the (saturated) source $\rho_\eta$ at $\z$.
The enclined dashed line represents the limit of the black disk, which 
increases linearly with $\eta$, as we shall see in Sect. 4.}
\label{long}
\end{center}
\end{figure}

There is a similar restriction on the values of $\eta$ :
For given $Q^2$, there is a minimum rapidity
$\bar\tau_1(Q^2)$ below which there is no black disk at all:
$R(\eta, Q^2)=0$ for $\eta<\bar\tau_1(Q^2)$.
This is the rapidity at which the black disk first emerges
at the center of the hadron, namely, at which
 $Q_s^2(\bar\tau_1,b=0) =Q^2$ (see Sect. 4.4
below). Thus, in order to count saturated sources
only, the integral over $\eta$ in Wilson lines like
(\ref{vtau}) must be restricted to the interval
$\bar\tau_1(Q^2) < \eta < \tau$. In Fig. \ref{long}, the saturated sources
(with momentum $Q^2$) occupy the lower right corner, below the 
dashed line which represents the profile of the black disk as a function
of $\eta$.

The external point $ x_\perp$ in eq.~(\ref{alpha1})
is at the impact parameter of the quark (or the antiquark) in the dipole,
so it satisfies $|x_\perp|\gg R(\tau, Q^2) \ge R(\eta, Q^2)$ for
any $\eta\le \tau$. It is therefore appropriate to
evaluate the field (\ref{alpha1}) in a multipolar expansion:
\be\label{alphaMP}
\alpha_{\eta}^{a}(x_\perp)&\!=\!&
\langle x_\perp|\frac{1}{-\grad^2_\perp}|0_\perp\rangle\int^R \!{d^2z_\perp}
\,\rho_{\eta}^a(z_\perp)- {\partial \over \partial x^i} 
\langle x_\perp|\frac{1}{-\grad^2_\perp}|0_\perp\rangle
\int^R \!{d^2z_\perp} z^i
\rho_{\eta}^a(z_\perp)\,+\cdots\nn
&\!\equiv \!&
\langle x_\perp|\frac{1}{-\grad^2_\perp}|0_\perp\rangle\ {\cal Q}^a
+ {x^i \over 2\pi x_\perp^2} \,{\cal D}^i_a\,+\cdots,\ee
where $R\equiv R(\eta, Q^2)$, 
$ {\cal Q}^a$ is the total colour charge within the
black disk, ${\cal D}^i_a$ is the corresponding dipolar moment, etc.

To compute the scattering amplitude (\ref{NS}),
one has to construct the Wilson lines $V^\dag$ and $V$ with the field
(\ref{alphaMP}) and then average over $\alpha$ 
(or, equivalently, over $\rho$) as
in eq.~(\ref{OBS}). In what follows, it is more convenient to
work with the probability distribution for $\rho$, i.e.,
$W_\tau[\rho]$. In general, this distribution is determined by a
complicated functional evolution equation, which is very non-linear
\cite{W,PI}.

However, as observed in Ref. \cite{SAT}, this equation
simplifies drastically in the saturation regime,  where the
corresponding solution $W_\tau[\rho]$ is essentially a Gaussian in $\rho$.
This can be understood as follows: 
The non-linear effects in the quantum evolution
enter via Wilson lines like eq.~(\ref{vtau}). At saturation, the
field $\alpha$ in the exponential carries typical
momenta  $\kk\sim Q_s$ and has a large amplitude
$\alpha \sim 1/g$. Thus, the Wilson lines are strongly varying
over a transverse distance $1/Q_s(\tau,\bb)$. When observed by a probe 
with transverse resolution $Q^2\ll Q_s^2(\tau,\bb)$,
these Wilson lines are rapidly oscillating and average to zero.
Thus, at saturation, one can drop out the Wilson lines,
and all the associated non-local and non-linear effects. Then,
the probability distribution
$W_\tau[\rho]$ becomes indeed a Gaussian, which by the same argument
is local in colour and space-time rapidity, and also homogeneous in
all the (longitudinal and transverse) coordinates. The only remaining
correlations are those in the transverse plane, and, importantly,
these are such as to ensure {\it colour neutrality} \cite{Cargese}.

Specifically, the only non-trivial correlation
function of the saturated sources is the two-point function,
which reads \cite{SAT} (see also Sect. 5.4 in Ref.  \cite{Cargese}) :
\be\label{rhoHM}
\langle \rho_{\eta}^{a}(z_\perp)\,
\rho_{{\eta}'}^{b}(u_\perp)\rangle_\tau&=&\delta^{ab}\delta({\eta}
-{\eta}')\,\lambda(z_\perp-u_\perp),\nn
\lambda(\kk)&=&\frac{1}{\pi}\,\kk^2.\label{rhok}\ee
For given $\eta$ and $\kk^2 \sim Q^2$,  eq.~(\ref{rhoHM}) holds
for points $z_\perp$ and $u_\perp$ within the black disk of radius
$R(\eta, Q^2)$. The crucial property of the 2-point 
function (\ref{rhoHM}) is that it vanishes as $\kk^2\to 0$. Physically,
this means that, globally, the saturated gluons 
are colour neutral\footnote{Since the distribution of $\rho$ is
a Gaussian, the fact that $\langle {\cal Q}^a{\cal Q}^a \rangle$ vanishes
is equivalent to ${\cal Q}^a=0$, which means colour neutrality indeed.},
as anticipated: $\langle {\cal Q}^a{\cal Q}^a \rangle=0$,
where ${\cal Q}^a$ is the total colour charge (at given $\eta$)
in the transverse plane.
In fact, since the Wilson lines average to zero over distances
$\Delta b_\perp \simge 1/Q_s(\eta,\bb)$, it follows that
colour neutrality is achieved already over a transverse scale of the order
of the saturation length:
\be
\int_{\Delta S_\perp} \!{d^2x_\perp} 
\int_{\Delta S_\perp} \!{d^2y_\perp}\, \langle\rho_{\eta}^{a}(x_\perp)
\rho_{\eta}^{a}(y_\perp)\rangle\,=\,0,\ee
where $\Delta S_\perp$ is, e.g., a disk of radius $R > 1/Q_s(\eta,\bb)$
centered at $\bb$.

This immediately implies that, as soon as the black disk is large
enough, the overall charge of the saturated gluons vanishes,
${\cal Q}^a=0$, so we can ignore the monopole field in eq.~(\ref{alphaMP}). 
Here, ``large enough'' means, e.g., $R(\tau, Q^2) \gg 1/Q$, which
guarantees that the radius of the black disk is larger than
the saturation length $1/Q_s(\eta,\bb)$ at any $\bb$ within the
disk and at any $\eta$ in the interval $\bar\tau_1(Q^2)
< \eta < \tau$ (since then $Q_s(\eta,\bb) >Q$).
In these conditions, the dominant field of the saturated gluons
at large distances is the {\it dipolar} field in eq.~(\ref{alphaMP}).

Below, we shall need the two-point function of this field:
\be\label{alphaHM}
\langle \alpha_\eta^{a}(x_\perp)\,
\alpha_{\eta'}^{b}(y_\perp)\rangle_\tau&\equiv &\delta^{ab}\delta(\eta
-\eta')\,\gamma_\eta(x_\perp-y_\perp).\ee
 From eq.~(\ref{alphaMP}), we deduce (with colour indices omitted,
since trivial):
\be
\gamma_\eta(x_\perp-y_\perp)\,=\,
{x^i \over 2\pi x_\perp^2} 
{y^j \over 2\pi y_\perp^2}\,\langle {\cal D}^i{\cal D}^j \rangle,\ee
with (cf. eq.~(\ref{rhoHM})):
\be
\langle {\cal D}^i{\cal D}^j \rangle&\equiv &
\int^R {d^2z_\perp} \int^R {d^2u_\perp}\ z^i u^j\
\lambda(z_\perp-u_\perp)\nn
&=&\int^R {d^2z_\perp} \int^R {d^2u_\perp}\ 
  \int^{Q_s} {d^2k_\perp \over (2\pi)^2}\
{\rm e}^{i k_\perp \cdot (z_\perp-u_\perp)}
{\partial^2 \over \partial k^i \partial k^j}\,\frac{\kk^2}{\pi}\nn
&\simeq&\delta^{ij} \,2 R^2(\eta,Q^2)\,.
\ee
where  $R\equiv R(\eta,Q^2)$, $Q_s\equiv Q_s(\eta,\bb)$ (with
$\bb \equiv (z_\perp+u_\perp)/2 < R$), and
formal manipulations like integrations by parts or the use of
the Fourier representation of the $\delta$-function were permitted
since $RQ_s \gg 1$. Thus, finally,
\be\label{gammaSAT}
\gamma_{\eta}(x_\perp,y_\perp)=
{1\over 2\pi^2}\ \frac{{x_\perp} \cdot{y_\perp}}{x_\perp^2 y_\perp^2}\
R^2({\eta},Q^2) ,
\ee
which, we recall, is valid only as long as $x_\perp,\,y_\perp
\gg R({\eta},Q^2)$.

We are now in a position to compute the scattering 
amplitude (\ref{NS}) for the  scattering between the incoming dipole
and the dipolar colour charge distribution within the black disk.
To this aim, we have to average the product 
${\rm tr}\big(V^\dagger(x_{\perp}) V(y_{\perp})\big)$ 
over the Gaussian random variable $\alpha_\eta$ with two-point
function (\ref{alphaHM}). The result of this calculation
is well-known (see, e.g., \cite{Cargese}):
\be\label{Stau2}
\tilde {\cal S}_\tau(x_\perp,y_\perp)= {\rm exp}\left\{
-{g^2C_F \over 2}
\int_{\bar\tau}^{\tau}d{\eta}
\Big[\gamma_{\eta}(x_{\perp},x_\perp)
+ \gamma_{\eta}(y_{\perp},y_\perp) - 2\gamma_{\eta}(x_{\perp},y_\perp)\Big]
\right\}
\ee
where $C_F \equiv t^at^a =(N_c^2-1)/2N_c$ and 
the lower limit $\bar\tau$ in the integral is a shorthand for
$\bar\tau_1(Q^2)$ with $Q^2=1/\rr^2$.
The ``tilde'' symbol on ${\cal S}$ is to remind that this is not the
total $S$-matrix element, but just the particular contribution
to it coming from the saturated gluons.
An immediate calculation using eqs.~(\ref{gammaSAT}) and (\ref{Stau2})
yields (with $\bar\alpha_s \equiv 2\alpha_s C_F/\pi \simeq \alpha_s N_c/\pi$
in the large $N_c$ limit) :
\be\label{Sdipole}
\tilde {\cal S}_\tau(x_\perp,y_\perp) &=&
{\rm exp}
\left\{-
\bar\alpha_s\ \frac{(x_\perp - y_\perp)^2}{2 x_\perp^2 y_\perp^2}\
\int_{\bar\tau}^{\tau}d{\eta}\ R^2({\eta},Q^2)
\right\}\nn
&\simeq&
{\rm exp}
\left\{-
\bar\alpha_s\ \frac{r_\perp^2}{2 b_\perp^4}\
\int_{\bar\tau}^{\tau}d{\eta}\ R^2({\eta},Q^2)
\right\},
\ee
where in the second line we have replaced in the denominator
$x_\perp^2\simeq y_\perp^2\simeq b_\perp^2$ (which is appropriate
since $b_\perp \gg R({\tau},Q^2) \gg \rr$). The exponent in
eq.~(\ref{Sdipole}) vanishes when the dipole shrinks to
a point, $\rr\to 0$. This is the expected dipole cancellation,
manifest already on eq.~(\ref{Stau2}).

The previous derivation makes the
 physical interpretation of eq.~(\ref{Sdipole})
very clear: The exponent there is the square of the potential
$\sim g t^a r_\perp \frac{1}{b^2} R(\eta)$ for the 
interaction between two dipoles --- the
``external dipole'' of size $r_\perp$ and the dipole made of the 
saturated gluons (at a given space-time rapidity $\eta$), 
with size $R(\eta)$ --- separated by a large distance $b$.
 There exists one layer of
saturated gluons at any $\eta$ within the interval
$\bar\tau_1(Q^2) < \eta < \tau$, so  eq.~(\ref{Sdipole})
involves an integral over this interval.

Of course, eq.~(\ref{Sdipole}) can be also obtained directly from the BK 
equation, although, in that context, its physical interpretation 
in terms of dipole--dipole scattering may not be so obvious. 
In fact, this is just a particular piece of what we have
called ``the contribution (B)'' in the previous subsection,
i.e., the contribution of the points $\z$ satisfying
$|\z-\bb| \gg 1/Q_s(\tau,\bb) $. If $\z$ now refers to the saturated
gluons within the black disk, then it is further restricted
by $|\z| < R(\tau,Q^2) $, which implies $|\z-\bb|\approx b$ for the same
reasons as above. Then, a simple calculation similar to
eq.~(\ref{IIcon0}) immediately yields
\BQA\label{SBK}
\frac{\del}{\del\tau}\tilde{\cal S}_\tau(x_\perp, y_\perp) 
 &\simeq & -\bar\alpha_s\, \frac{\rr^2}{\bb^4} \, 
R^2(\tau,Q^2)\, \tilde{\cal S}_\tau(x_\perp, y_\perp),
\EQA
which  after integration over $\tau$ is indeed equivalent to 
eq.~(\ref{Sdipole}) \footnote{The mismatch by a factor of two between
eqs.~(\ref{Sdipole}) and (\ref{SBK}) is inherent to the mean field 
approximation used in Refs. \cite{SAT,Cargese} to derive eq.~(\ref{Stau2}),
and is completely irrelevant for the kind of estimates that we are
currently interested in.}. 

For comparison with eq.~(\ref{Sdipole}), it is interesting to compute
also the $S$-matrix element for a {\it coloured} external probe, 
e.g., a quark, which scatters off the saturated gluons
in the eikonal approximation.
A calculation entirely similar to that leading to eq.~(\ref{Sdipole})
yields ($\bb$ is the transverse location of the quark):
\be\label{SKW}
\frac{1}{N_c}\,
\Big\langle {\rm tr} \,V^\dagger(b_{\perp}) 
\Big\rangle_\tau&=&{\rm exp}\left\{
-{g^2 C_F \over 2}
\int_{\bar\tau}^{\tau}d{\eta}
\ \gamma_{\eta}(b_{\perp},b_\perp)\right\}\nn
&\simeq& 
{\rm exp}
\left\{-
\bar\alpha_s\ \frac{1}{2b_\perp^2}\
\int_{\bar\tau}^{\tau}d{\eta}\ R^2({\eta},Q^2)
\right\}
\ee
where the second line follows after using eq.~(\ref{gammaSAT}).

To summarize,
the amplitude for the scattering off the saturated gluons decreases
like $1/b_\perp^4$ for an external dipole, but only as $1/b_\perp^2$ for
a coloured probe. This difference turns out to be essential: because
of it, this long-range
scattering plays only a marginal role for the dipole, while it
leads to unitarity violations in the case of the coloured probe
(although the very question
of unitarization makes little physical sense
for a ``probe'' which is not a colour singlet).

To see this, assume the  long-range contributions shown above
to be the {\it only} contributions, or, in any case, those which
give the dominant contribution to the cross-section. Then, one can rely
on the previous formulae to estimate the rate of expansion of the black
disk. Namely, assume that, for the purpose of getting
an order-of-magnitude estimate, one can extrapolate 
eqs.~(\ref{Sdipole}) and (\ref{SKW}) up to energies where
the black disk approaches the incidence point $\bb$
of the external probe. Then, we expect
the exponents in these equations to become of order
one for $b\sim R({\tau},Q^2)$. For the external dipole, this 
condition implies:
\be\label{sat-est}
\bar\alpha_s\ \frac{r_\perp^2}{R^4(\tau, Q^2)}
\int_{\bar\tau}^{\tau}d{\eta}\ R^2({\eta},Q^2)\,\sim\,1.\ee
This gives (recall that $Q^2= {1 /r_\perp^2}$):
\be
Q^4 R^4(\tau, Q^2)\,=\,\bar\alpha_s \int_{\bar\tau}^{\tau}d{\eta}\ 
Q^2R^2({\eta},Q^2),\ee
or, after taking a derivative w.r.t. $\tau\,$,
\be\label{Rtau}
2\,\frac{d}{d\tau} \,\Big(Q^2 R^2(\tau, Q^2)\Big)\,=\,
\bar\alpha_s,\ee
whose solution $R^2(\tau, Q^2)$ increases linearly with
$\tau$. 

By contrast, for a coloured probe, the same condition
yields:
\be \label{Rtau-KW0}
\bar\alpha_s
\int_{\bar\tau}^{\tau}d{\eta}\ R^2({\eta},Q^2)\,=\,{R^2(\tau, Q^2)}\,\ee
or after taking a derivative w.r.t. $\tau$:
\be\label{Rtau-KW}
\frac{d}{d\tau} \,R^2(\tau, Q^2)\,=\,
\bar\alpha_s\,R^2(\tau, Q^2),\ee
which gives an exponential increase with $\tau$, as found in
Ref. \cite{KW02}.

We thus see that the violation of unitarity by long-range Coulomb
scattering reported by the authors of Ref. \cite{KW02} is related
to their use of an external probe which carries a non-zero colour charge.
{ This case
is physically ill defined, and therefore uninteresting (note, indeed, that 
$\langle {\rm tr} \,V^\dagger \rangle$ is not a gauge-invariant
quantity); in particular, its relevance for the problem of gluon saturation 
in the target wavefunction 
remains unclear to us (since the relation between ``blackness''
and saturation holds only for dipole probes; cf. the discussion prior
to eq.~(\ref{INT_sat})).

On the other hand, for the physically interesting case of an 
external {\it dipole}, the contribution (\ref{Rtau}) to the expansion of
the black disk not only is consistent with unitarity
--- if this was the {\it only} contribution, the cross-section 
$\propto R^2(\tau, Q^2)$ would increase { linearly} with $\tau$ ---, but
at large $\tau$, is even negligible as compared to the corresponding
contribution of the short-range scattering 
(which gives a cross-section increasing like $\tau^2$, cf. 
eqs.~(\ref{BDR})--(\ref{sigmaFC})).

These considerations are conveniently summarized in the following,
schematic, approximation to the scattering amplitude in the grey area,
which follows from the previous analysis in this section:
${\cal N}_\tau(\rr,\bb)$ is the sum of two contributions,
a short-range contribution, cf.
eq.~(\ref{Nfact1}), and a long-range one, cf. eq.~(\ref{Sdipole}) :
\be\label{Ngrey}
{\cal N}_\tau(\rr,\bb)\,\approx\,
 \sqrt{\rr^2\Lambda^2}\,
{\rm e}^{\omega\bar\alpha_s\tau}\,{\rm e}^{-2m_{\pi}b}\,+\,
\bar\alpha_s\,\frac{r_\perp^2}{2 b_\perp^4}
\int_{\tau_0}^{\tau}d{\eta}\ R^2({\eta},Q^2)\,,\ee
with the short-range contribution determined by the solution
to the homogeneous BFKL equation (\ref{BFKLb}) together with the
assumed exponential fall-off of the initial condition
(see Sects. 4 and 5 below for more details),
and the long-range contribution obtained by keeping only the lowest-order term
in eq.~(\ref{Sdipole}) (which is enough since we are in a
regime where ${\cal S}_\tau(x_\perp,y_\perp)\sim 1$). At
the initial rapidity $\tau_0$, the long-range contribution vanishes,
while the short-range contribution reduces to
${\rm e}^{-2m_{\pi}(b-R(\tau_0,Q^2))}$ with $R(\tau_0,Q^2)$ given by
eq.~(\ref{BDR}), as it should\footnote{That is, the initial
scattering amplitude ${\cal N}_{\tau_0}(\rr,\bb)$ is equal to one
within the black disk ($\bb \le R(\tau_0,Q^2)$), and decreases exponentially
outside it. Eq.~(\ref{Ngrey}) applies, of course, only at impact parameters
outside the black disk.}.

 For given $\tau$ and $\rr= 1/Q$, eq.~(\ref{Ngrey}) applies
at impact parameters in the grey area, $R(\tau,Q^2)< b < R_H(\tau)$, but it can
be extrapolated to estimate the boundaries of this area,
according to eqs.~(\ref{INT_sat})--(\ref{RH}).
It is easy to check that, at high energy, both these boundaries
are determined by the { short-range} contribution, i.e., the first term
in the r.h.s. of eq.~(\ref{Ngrey}). Thus, this contribution dominates the
scattering amplitude at any $\bb$ in the grey area. By comparison,
the long-range contribution is suppressed by 
one power of $1/\bar\alpha_s\tau$.

For instance, the black disk radius is obtained by requiring
${\cal N}_\tau(\rr,\bb)\sim 1$ for $\bb\sim R(\tau,Q^2)$ 
(cf. eq.~(\ref{INT_sat})). If one assumes the short-range contribution
to dominate in this regime, one obtains the estimate (\ref{BDR}) for
the black disk radius: $
R(\tau,Q^2)\simeq (\omega/2m_\pi)\bar\alpha_s \tau$. By using this result,
one can  evaluate the corresponding long-range contribution, and thus
check that this is comparatively small, 
as it should for consistency with the original assumption:
\be
\bar\alpha_s\ \frac{r_\perp^2}{R^4(\tau, Q^2)}
\int^{\tau}d{\eta}\ R^2({\eta},Q^2)\,\bigg
|_{R(\tau,Q^2)\sim \,\,\bar\alpha_s \tau/m_\pi}\,\,\sim\,\,
\frac{\rr^2 m_{\pi}^2}{\bar\alpha_s\tau}\,\,\ll\,\,1.\ee
By contrast, if one starts by assuming that the long-range contribution
dominates, then one is running into a contradiction,
since in this case $R(\tau,Q^2)\propto \sqrt{\bar\alpha_s\tau}$,
cf. eq.~(\ref{Rtau}), and the short-range contribution increases
exponentially along the ``trajectory'' $b= R(\tau,Q^2)$.

A similar conclusion holds for $b=R_H(\tau)$
[since $R_H(\tau) = R(\tau,Q^2=\Lambda^2_{QCD}) $, cf. eq.~(\ref{RH})],
and therefore for any point $\bb$ within the grey area. (The hadron
radius $R_H(\tau)$ will be evaluated in Sect. 4.2 below.) Thus, the
short-range contribution is indeed the dominant one in the grey area.
This contribution preserves the exponential fall-off of the initial condition,
and therefore saturates the Froissart bound, as explained in Sect. 2.

It is essential for the consistency of the previous arguments --- 
which combine perturbative quantum evolution with non-perturbative 
initial conditions --- that perturbation theory has been applied only 
in the regime where it is expected to be valid, namely,
in the central region at $b<R_H(\tau)$, where the gluon density
is high and the local
saturation scale $Q_s(\tau,b)$ is much larger than $\Lambda_{QCD}$.
It has been enough to consider this region for the 
present purposes since this includes both the black disk and the grey area
which controls its expansion. Within this region, the perturbative
evolution equations of Refs. \cite{B,K,W,PI} can be
trusted, and the additional approximations that we have performed
on these equations are under control as well. When supplemented with
appropriate boundary conditions --- which are 
truly non-perturbative, since reflecting the physics of confinement
---, these equations allow one to compute the
rate for the expansion of the black disk, and, more generally, to follow
this expansion as long as the black disk remains confined within the
region of applicability of perturbation theory (which includes, at least,
the central area at the initial rapidity $\tau_0$ : $b<R_H(\tau_0)$).

But if one attempts to follow this expansion up to much higher
rapidities, where $R(\tau,Q^2) \gg R_H(\tau_0)$, then
a strict application of the perturbative evolution {\it with initial
conditions at} $\tau=\tau_0$ would run into 
difficulties\footnote{The discussion in this paragraph has been inserted
 as a partial response to criticism by Kovner and
Wiedemann \cite{KW022}, written in response to the original version of
this paper.  Please see the note added at the end of the paper for further
discussion.}. The difficulties
arise since, in the absence of confinement, the long-range dipolar tails
created by the saturated gluons can extend to arbitrary large distances,
and thus contribute to scattering even at very large impact parameters
($b \gg R_H(\tau_0)$), where physically there should be no contribution at all.
This is illustrated by eq.~(\ref{Ngrey}): We have previously argued that,
for $\bb$ within the grey area, the dominant contribution 
comes from short-range scattering 
(the first term in the r.h.s. of eq.~(\ref{Ngrey})). But if one
extrapolates this formula at very large $b \gg R_H(\tau)$, then, clearly,
the long-range contribution, which has 
only a power-law fall-off with $\bb$, 
will eventually dominate over the short-range 
contribution, which decreases exponentially. When this happens, however,
the impact parameters are so large ($b-R(\tau,Q^2) \gg 1/m_\pi$) that the
long-range scattering is controlled by the exchange of very soft ($\kk\simle
m_\pi$) quanta, which in a full theory would be suppressed by the confinement.
That is, in a more complete theory which would include the physics of
confinement, the long-range contribution to eq.~(\ref{Ngrey}) would be
suppressed at very large $\bb$ by an additional factor
${\rm e}^{-2m_{\pi}(b-R(\tau,Q^2))}$, so that the short-range contribution
will always dominate, for {\it all} impact parameters.
But in the present, perturbative, setting,
the only way to avoid unphysical long-range
contributions is to start the quantum evolution directly in the grey area,
as we did before 
 (rather than try to construct this grey area via 
perturbative evolution from earlier rapidities $\tau_0\ll \tau$,
at which the points $\bb$ of interest were far outside the {\it initial}
grey area: $b \gg R_H(\tau_0)$). 

Note finally that what is truly remarkable, and also essential for our
conclusion on the Froissart bound, is not the suppression of the long-range
{\it non-perturbative} contributions by the confinement
--- this is only to be expected in the full theory, 
and can be also enforced in the present calculation
by appropriately chosing the boundary conditions ---, but rather the suppression
of the long-range {\it perturbative} contribution {\it within the grey area}
(where perturbation theory applies, so its predictions must be taken
at face value). We mean here, of course, the fact that the long-range
contribution to eq.~(\ref{Ngrey}) falls off like 
$1/\bb^4$, and not like $1/\bb^2$, as it would have been the case if the
saturated gluons were uncorrelated. The mechanism for this suppression
is purely perturbative, and related to saturation: The saturated gluons
are globally colour neutral, so the monopole fields of the individual
gluons are replaced at large distances $\gg 1/Q_s$ by the more
rapidly decreasing dipolar field of the whole distribution.
To understand the relevance of this suppression for the Froissart bound,
consider
what would happen if the saturated sources were {\it statistically independent}, 
i.e., if eq.~(\ref{rhoHM}) was replaced by
$$ \langle \rho_{\eta}^{a}(z_\perp)\,
\rho_{{\eta}'}^{b}(u_\perp)\rangle_\tau\,=\,\lambda\,\delta^{ab}\delta({\eta}
-{\eta}')\delta^{(2)}(z_\perp-u_\perp)\,.$$
Then the  exponent in eq.~(\ref{Sdipole}), and also the second term
in the r.h.s. of eq.~(\ref{Ngrey}), would change into
$$\bar\alpha_s\, \frac{\lambda r_\perp^2}{{b_\perp^2}}\,
\int^\tau_{\bar\tau} d\eta \,R^2({\eta},Q^2)\,,$$
which would generate a black disk increasing exponentially
with $\tau$ (cf. eq.~(\ref{Rtau-KW0})--(\ref{Rtau-KW})). Thus, the
long-range contribution would dominate already within the grey area,
and the Froissart bound would be violated. We thus conclude that
colour correlations at saturation are essential to ensure unitarity.

}

\section{Black disk evolution and the \FB}\label{BDFB}
\setcounter{equation}{0}

In this section, we shall exploit the factorization property (\ref{Nfact})
together with the known solution ${\cal N}_\tau(\rr)$ to the homogeneous
BFKL equation in order to compute the scattering amplitude in the grey area,
and thus study the evolution of the black disk with increasing energy. 
After briefly recalling the BFKL solution, in Sect.~4.1, we shall then
compute the radius of the black disk $R(\tau,Q^2)$ and derive the 
\FB (in Sect.~4.2). Then, in Sect.~4.3, we shall study the impact parameter
dependence of the saturation scale $Q_s^2(\tau,\bb)$, and deduce a physical picture 
for the expansion of the black disk, to be exposed in Sect.~4.4.

\subsection{Scattering amplitude in the BFKL approximation}

Eq.~(\ref{Nfact}) for the scattering amplitude in the grey area
involves the solution ${\cal N}_\tau(\rr)$
to the homogeneous BFKL equation, i.e., 
the BFKL equation without impact parameter dependence. 
This solution is well known, and we shall briefly recall here the relevant
formulae, at the level of accuracy of the present calculation.
(See Refs. \cite{AKM,SCALING} for a similar approach and more details.)

The solution can be expressed as a Mellin transform 
with respect to the transverse coordinate:
\BQ\label{Mellin}
{\cal N}_\tau(\rr=1/Q)=\int_{C} \frac{d\lambda}{2\pi i} 
\left(\frac{\Lambda^2}{Q^2}\right)^\lambda  {\rm e}^{\bar\alpha_s\tau \{
2\psi(1)-\psi(\lambda)-\psi(1-\lambda)
\}}\, ,
\EQ
where $\psi(\lambda)$ is the di-gamma function, and $\Lambda$ is an
arbitrary reference scale, of order $\Lambda_{QCD}$. The contour $C$
in the inverse Mellin transform
is taken on the {left} of all the singularities of the integrand
in the half plane Re $\lambda>0$. Note that, since ${\cal N}_\tau(\rr)$
is a function of $\rr^2$, we find it convenient to
use the momentum variable $Q^2=1/\rr^2$ to characterize the transverse
resolution of the dipole. From now on, we shall again use the notation 
${\cal N}_\tau(Q^2)\equiv {\cal N}_\tau(\rr=1/Q)$, which was already
introduced in Sect.~2 (cf. eq.~(\ref{INT_sat})).

We are interested here in a regime where the energy is very  high,
$\bar\alpha_s\tau\gg 1$, and the dipole is small: $Q^2\gg \Lambda^2$.
In these conditions, it is appropriate to evaluate the
integral (\ref{Mellin}) in the saddle point approximation. Higher
is the energy, better is justified this approximation, and closer is the
saddle point $\lambda_0$ --- which is a function of
$(\ln Q^2/\Lambda^2)/\bar\alpha_s\tau$ --- of the so-called ``genuine
BFKL'' saddle-point at $\lambda_0=1/2$. (This is the saddle point
which governs the asymptotic behaviour of the solution to the
BFKL equation at very large energy.) In fact, for
\BQ\label{r/t}
\frac{1}{\bar\alpha_s\tau}\ln \frac{Q^2}{\Lambda^2}\,\ll \,1,
\EQ
which is the most interesting regime here, the saddle point is
easily estimated as:
\be\label{BFKL_SP}
\lambda_0\,\simeq \,\frac{1}{2} +
\frac{1}{\beta\bar\alpha_s\tau}\ln \frac{Q^2}{\Lambda^2}\,,\ee
with $\beta=28\zeta(3)$. In fact,
the recent analysis in Ref. \cite{SCALING} shows that eq.~(\ref{BFKL_SP})
remains a good approximation for the saddle point even for comparatively
low energies, such that $\bar\alpha_s\tau\sim \ln (Q^2/\Lambda^2)$.
This saddle point gives the standard BFKL solution, which, after
multiplication with the profile function (cf. eq.~(\ref{Nfact})), 
provides the 
scattering amplitude in the grey area in the present approximation:
\BQ\label{BFKL_sol}
{\cal N}_\tau(Q^2, \bb)\,\simeq\, S(\bb)\, \sqrt{\frac{\Lambda^2}{Q^2}}
\,{{\rm e}^{\omega\bar\alpha_s\tau}\over\sqrt{2\pi \beta \bar\alpha_s \tau} }
\, \exp\left\{
-\frac{1}{2\beta\bar\alpha_s\tau}\left(\ln \frac{Q^2}{\Lambda^2}\right)^2
\right\},
\EQ 
where $\omega=4\ln 2$ is the customary BFKL exponent.
The factor $\sqrt{2\pi \beta \bar\alpha_s \tau}$ in the denominator
comes from integrating over the Gaussian fluctuations around the 
saddle point. When exponentiated, this gives a contribution
$\propto \ln(\bar\alpha_s\tau)$ which is subleading at large
energy and will be ignored in what follows. It is then convenient
to rewrite eq.~(\ref{BFKL_sol}) as follows:
\BQ\label{BFKL_sol1}
{\cal N}_\tau(Q^2,\, \bb)\,\simeq\,
\exp\left\{
-2m_\pi b + \omega\bar\alpha_s\tau - \frac12\ln \frac{Q^2}{\Lambda^2}
-\frac{1}{2\beta\bar\alpha_s\tau}\left(\ln \frac{Q^2}{\Lambda^2}\right)^2
\right\},
\EQ
where we have also used $S(\bb)\approx {\rm e}^{-2m_{\pi}b}$, as appropriate
for sufficiently large $b$ ($b \gg R_0$, cf. the discussion after eq.~(\ref{WS})).
This is the most interesting case here, since we consider the high energy
regime in which the black disk is already quite large. 

Eq.~(\ref{BFKL_sol1}) is valid for those values of the
parameters $\tau$, $Q^2$ and $\bb$ for which our previous approximations
are justified, namely, such that the conditions
$\Lambda^2\ll Q_s^2(\tau,b)\ll Q^2$ are satisfied.
As it was anticipated in Sect. 2, and will be verified below in this section,
these conditions are realized within a corona at 
$R(\tau,Q^2) \ll \bb \ll R_H(\tau)$, which, with increasing energy,
moves further and further away from the center of the hadron.

When decreasing $\bb$ towards $R(\tau,Q^2)$ at fixed $\tau$, or,
equivalently, increasing $\tau$ at fixed $\bb$, the scattering amplitude
increases towards one, and the BFKL approximation (\ref{BFKL_sol1}) ceases
to be valid. (The dipole resolution $Q^2$ is always fixed in these
considerations.) But it is nevertheless legitimate to use eq.~(\ref{BFKL_sol1}) 
in order to estimate the boundary of its range
of validity, that is, the black disk radius $R(\tau,Q^2)$, or the 
saturation scale $Q_s^2(\tau,\bb)$. Indeed, the non-linear effects become
important when the BFKL solution (\ref{BFKL_sol1}) becomes of order one.
This condition can be written
either as an equation for $R(\tau,Q^2)$ for given $\tau$ and $Q^2$, 
namely, eq.~(\ref{INT_sat}),  
or as an equation for $Q_s^2(\tau,\bb)$
for given $\tau$ and $\bb$, namely, eq.~(\ref{INT_sat1}). 
(One could, of course, similarly introduce and compute also a
critical rapidity $\bar\tau(Q^2,\, \bb)$ at which blackness is reached
for given $Q^2$ and $\bb$, but this is less interesting for our
subsequent discussion. See, however, Sect. 4.4.)

\subsection{The black disk radius}

In this subsection, we shall use eqs.~(\ref{INT_sat}) and (\ref{BFKL_sol1})
to compute the radius of the black disk and study some limiting cases.
Eq.~(\ref{INT_sat}) amounts to the condition that the exponent
in eq.~(\ref{BFKL_sol1}) 
vanishes\footnote{At the level of the present approximation, one can take 
$\kappa=1$ in eqs.~(\ref{INT_sat}) and (\ref{INT_sat1}) without loss of 
accuracy.}, which immediately implies:
\BQ\label{BDR_full}
2m_\pi R(\tau,Q^2)=\omega \bar\alpha_s \tau -\frac12 \ln \frac{Q^2}{\Lambda^2}
-\frac{1}{2\beta\bar\alpha_s\tau}\left(\ln \frac{Q^2}{\Lambda^2}\right)^2.
\EQ
The right hand side is positive as long as $Q^2 <
 \Lambda^2\, {\rm e}^{c\bar\alpha_s\tau}\equiv Q_s^2(\tau,b=0)$, with
\be\label{c}
c\equiv -\frac{\beta}{2}+ \frac{1}{2}\sqrt{\beta(\beta+8\omega)}\,=\,
4.84...\,.\ee
As anticipated by our notations,
\be\label{Qs0}
Q_s^2(\tau,b=0)\,=\,\Lambda^2 
{\rm e}^{c\bar\alpha_s \tau}\ee
is the saturation scale at the center of the hadron (this will be
verified via a direct computation in the next subsection). This is as expected: for
$Q^2 \ge Q_s^2(\tau,b=0)$, the hadron looks grey everywhere, so $R(\tau,Q^2)=0$.

The other extreme situation is when $Q^2\simeq \Lambda^2$, so that the black
disk extends up to the edge of the hadron (cf. eq.~(\ref{RH})).
Equation~(\ref{BDR_full}) yields then:
\be\label{RH1}
R_H(\tau)\,\approx\,\frac{\omega \bar\alpha_s }{2m_\pi}\, \tau \, ,\ee
which should be seen  only as a crude estimate: for such a small $Q^2$,
our approximations are not justified any longer.

But the physically interesting case is when $Q^2\gg \Lambda^2$,
but the energy is so large that the condition (\ref{r/t}) is satisfied.
Then one can neglect the term quadratic in $\ln Q^2/\Lambda^2$
in eq.~(\ref{BDR_full}) (since this term vanishes when $\tau\to \infty$),
and deduce that:
\BQ\label{BD_asy}
R(\tau,Q^2)\, \simeq \, \frac{1}{2m_\pi}\left( \omega \bar\alpha_s \tau 
-\frac12 \ln \frac{Q^2}{\Lambda^2}\right).
\EQ
The term linear in $\ln Q^2/\Lambda^2$, although subleading at large $\tau$
(since independent of $\tau$), has been nevertheless kept in the above
equation
since, first, we expect this term to give the dominant $Q^2$--dependence
of the cross-section at high energy, and, second, it measures the 
separation between the black disk and the edge of the hadron in the high
energy regime. Specifically:
\beq\label{RR}
R_H(\tau)\,-\,R(\tau,Q^2)\,\approx\,\frac{1}{4m_\pi}\, 
\ln \frac{Q^2}{\Lambda^2}\,,
\ee
which is fixed (i.e., independent of $\tau$), 
but large when $Q^2\gg \Lambda^2$. This is important since the points
$\bb$ at which our approximations are justified
should lie deeply within this corona: $R(\tau,Q^2) \ll b \ll R_H(\tau)$.
Thus, as anticipated in Sect. 2, the fact that $Q^2\gg \Lambda^2$
ensures the existence of a large grey area in which our
approximations apply.

Equation~(\ref{BD_asy}) is our main result in this paper. It
shows that, at  very high energy,
the radius of the black disk increases only
linearly with $\tau$, i.e., logarithmically with the energy.
This is the result anticipated in eq.~(\ref{BDR}). The corresponding
cross-section is given by eq.~(\ref{sigmaFC}) and saturates the \FBB,
that is, it grows like $\ln^2 s$, with a proportionality coefficient
which is {\it universal} (i.e., the same for any hadronic target), and
which reflects the combined role of perturbative and
non-perturbative physics in controlling the asymptotic behaviour
 at high energy.

In the remaining part of this paper, we shall further explore
this result and gain a different perspective over it by computing
also the saturation scale and studying the geometric scaling properties.

\subsection{Saturation scale with the impact parameter dependence}

In this subsection, we shall compute the saturation scale
for an inhomogeneous hadron and study its variation with the energy
and the impact parameter. Previous studies of this kind were restricted
to a homogeneous hadron \cite{AM2,SCALING,MuellerT02}, but, as
we shall see, the dependence upon the impact parameter introduces
some interesting new features. 

By inspection of eq.~(\ref{BFKL_sol1}), it is clear
that the saturation condition (\ref{INT_sat1}) amounts to the following,
second order algebraic equation for the quantity  
$\rho_s\equiv(1/\bar\alpha_s\tau)\ln Q_s^2/\Lambda^2$:
\BQ\label{2nd_order_eq}
\rho_s^2 + \beta \rho_s -2\beta \omega 
=-2\beta \frac{2m_\pi b}{\bar\alpha_s\tau}\,.
\EQ
The solution to this equation and the corresponding
saturation scale read:
\BQA
\rho_s(\tau,b)&=&-\frac{\beta}{2}+ 
\frac{\beta}{2}\,\sqrt{1+
\frac{8\omega} {\beta}\left(1-\frac{2m_\pi b}
{\omega\bar\alpha_s\tau}\right)}\,,
\label{rho_s_exact}\\
Q_s^2(\tau,b)&=& \Lambda^2\, {\rm e}^{\bar\alpha_s\tau\rho_s(\tau,b)}. 
\label{Qs_exact}
\EQA
Note that, in general, the 
impact parameter dependence in the saturation scale (\ref{Qs_exact}) is 
not factorizable. Below, however, we shall recover factorization
in some specific limits.

It can be easily checked that the above equations 
(\ref{rho_s_exact})--(\ref{Qs_exact}) and eq.~(\ref{BDR_full})
are consistent with each other, in
the sense that $Q_s^2(\tau,b=R(\tau,Q^2))=Q^2$, as it should
(cf. eq.~(\ref{condR})).
In particular, one can use eqs.~(\ref{rho_s_exact})--(\ref{Qs_exact})
to rederive the results in eqs.~(\ref{c})--(\ref{Qs0}) for the
saturation scale $Q_s^2(\tau,b=0)$ at the center of the hadron, as well
as eq.~(\ref{RH1}) for the hadron radius.
\begin{figure}
\begin{center}
\includegraphics[width=.8\textwidth]{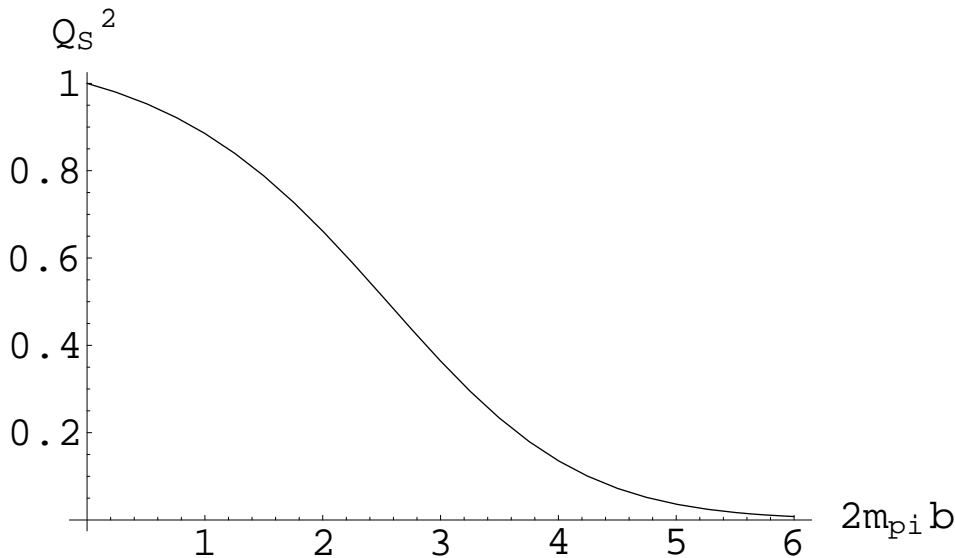}
\caption{The saturation scale $Q_s^2(b)/Q_s^2(b=0)$ 
from eqs.~(\ref{rho_s_exact})--(\ref{Qs_exact}) for $\bar\alpha_s\tau
=3$ and the Woods-Saxon profile function of eqs.~(\ref{WS})--(\ref{Sb}) 
with $R_0=3/2m_\pi$. On the abscisa,
the radial distance is measured in units of $1/2m_\pi$.}
\label{Woods-Saxon}
\end{center}
\end{figure}

A pictorial representation of the $b$--dependence of the
saturation scale, 
as emerging from eqs.~(\ref{rho_s_exact})--(\ref{Qs_exact}),
is given in Fig. \ref{Woods-Saxon}. As compared
to eq.~(\ref{rho_s_exact}), in this graphical representation we have replaced
$2m_\pi b\longrightarrow -\ln S(b)$, 
with $S(b)$ given by a Woods-Saxon profile, cf. eqs.~(\ref{WS})--(\ref{Sb});
this is more realistic than the exponential at short distances,
$b\simle R_0$, where it has a much slower decrease, but it shows
the expected fall-off $S(b)\approx {\rm e}^{-2m_{\pi}(b-R_0)}$
at larger distances.
As manifest on this figure, $Q_s^2(\tau,b)$ is itself very similar to
an exponential for all distances  $b\simge R_0$. This can be 
understood via a further study of eq.~(\ref{rho_s_exact}), which
will also reveal that, in fact, there is a change in the slope of the
exponential with increasing $b$ : To a very good
approximation, the plot in Fig. \ref{Woods-Saxon} can be seen as
the superposition of two exponentials, one at small
$b$, the other one at large $b$, which have different exponential slopes.

To see this, note that the function 
\be\Delta(\tau,b)\,\equiv\,
 1\,-\,\frac{2m_\pi b}{\omega\bar\alpha_s\tau}\,=\,1\,-\,\frac{b}
{R_H(\tau)}\,\,,\ee
 which enters
the square root in eq.~(\ref{rho_s_exact}), is positive semi-definite
for $b$ within the hadron radius ($b\le R_H(\tau)$), and monotonically
decreasing with $b$ from $\Delta(\tau,b=0)=1$ to 
$\Delta(\tau,b=R_H(\tau))=0$. This suggest two different approximations
according to whether $\Delta$ is close to one (for $b$ sufficiently
small) or close to zero (for $b$ sufficiently close to $R_H(\tau)$).
(Note that the factor $8\omega/\beta$ multiplying
$\Delta(\tau,b)$ in eq.~(\ref{rho_s_exact}) is a number
of order one, $8\omega/\beta\approx 0.67$, so it does not interfere
with our order-of-magnitude estimates.)

\noindent{\bf (I)} 
If $\Delta$ is close to one, which happens when $b$ is much smaller
than the hadron radius:
\be\label{smallb}
\frac{2m_\pi b}{\omega\bar\alpha_s\tau}\,\ll\,1,\qquad {\rm or}\qquad
b\,\ll\,R_H(\tau),\ee
one can evaluate the square root in eq.~(\ref{rho_s_exact}) in an expansion
in powers of $1-\Delta$ (this is equivalent to an expansion of
$\rho_s(\tau,b)$ in powers of $b$
around $\rho_s(\tau,b=0)\equiv c$, cf. eq.~(\ref{c})). 
To linear order in this expansion, one obtains:
\be\label{rhosmall}
\rho_s(\tau,b)&\simeq&c\,-\,\frac{2m_\pi b}{\lambda_s\bar\alpha_s\tau}\,,
\qquad \lambda_s\equiv\sqrt{{ \beta+8\omega\over 4\beta}}
\,=\, 0.644...\,,
\ee
which gives (with $\gamma\equiv 1/\lambda_s\approx 1.55$):
\BQ\label{Qscenter}
Q_s^2(\tau,b)\simeq \Lambda^2\, {\rm e}^{c\bar\alpha_s\tau} \, 
{\rm e}^{-2\gamma m_\pi b}\,\equiv\,Q_s^2(\tau,b=0)\,[S(b)]^\gamma\, .
\EQ 
This is in a factorized form, although, as compared to the
corresponding factorized structure of the scattering amplitude
(\ref{Nfact}), it features some ``anomalous dimension'' $\gamma$
for the profile function. The value $Q_s^2(\tau,b=0)$ at
the center of the hadron is the same as the saturation scale
for a homogeneous hadron previously found in Refs. 
\cite{AM2,SCALING,MuellerT02}. Also, the constant $\lambda_s$
which appears in eq.~(\ref{rhosmall}) is the value of the
saddle point $\lambda_0$ in the Mellin representation
(\ref{Mellin}) for $Q^2=Q_s^2(\tau,b=0)$ (i.e., eq.~(\ref{BFKL_SP})
with $\ln (Q^2/\Lambda^2)=c\bar\alpha_s\tau$) \cite{SCALING}.

 Equation~(\ref{Qscenter}) shows that
the saturation scale decreases exponentially with the distance
$b$ from the center of the hadron,
with a typical decay scale $\sim 1/2\gamma m_\pi$.
(Of course, this exponential law applies
only for values $b$ which are not too close to the center, 
$b\simge R_0$, cf. Fig. \ref{Woods-Saxon}.)

\noindent{\bf (II)} 
When $b$ is sufficiently close to $R_H(\tau)$, in the sense that:
\be \label{largeb}
\Delta(\tau,b)\,\equiv\,\frac{R_H(\tau)-b}{R_H(\tau)}\,\ll\,1,\ee
than one can expand eq.~(\ref{rho_s_exact}) in powers of $\Delta$ (this
is an expansion  of $\rho_s(\tau,b)$ around $\rho_s(\tau,b=R_H) =0$).
To lowest order in this expansion, one obtains $\rho_s(\tau,b)\simeq
2\omega\Delta(\tau,b)$, and therefore:
\BQA
Q_s^2(\tau,b)\,\simeq \,\Lambda^2\, {\rm e}^{2\omega \bar\alpha_s \tau}
\, {\rm e}^{-4m_\pi b}.\label{Qs_asy}
\EQA
Thus, the saturation scale in the tail of the hadron distribution is still
in a factorized form, but the exponential slopes are different as compared
to the corresponding form near the center,  eq.~(\ref{Qscenter}),
both for the
increase with $\tau$ --- which is now controlled by the BFKL exponent
$2\omega\approx  5.55 > c$ --- and for the decrease with $b$ ---
where the ``anomalous dimension'' $\gamma$ of eq.~(\ref{Qscenter})
has been now replaced by 2.

These changes can be easily understood by reference to
eq.~(\ref{BFKL_sol1}): The exponent there must vanish when
$Q^2=Q_s^2(\tau,b)$. If $b$ satisfies the condition (\ref{largeb}),
then the first two terms in the exponent, which are the {\it large}
terms, almost cancel each other, so the other terms there
must be relatively small, in the sense
of eq.~(\ref{r/t}). Then, the term  quadratic in $\ln (Q^2/\Lambda^2)$
is much smaller than the linear term, and can be neglected. We thus end
up with
\BQ\label{BFKL_sol_asy}
{\cal N}_\tau(Q^2,\, \bb)\,\approx\,
\exp\left\{
-2m_\pi b + \omega\bar\alpha_s\tau - \frac12\ln \frac{\Lambda^2}{Q^2}
\right\}\,=\, \sqrt{\frac{\Lambda^2}{Q^2}}\,
{\rm e}^{\omega \bar\alpha_s \tau-2m_\pi b},\EQ
which, together with the saturation criterion (\ref{INT_sat1}),
provides indeed the expression  (\ref{Qs_asy}) for the saturation scale.
To summarize, when the ``diffusion'' term in the BFKL solution 
(\ref{BFKL_sol}) becomes negligible, then the energy dependence
and the $Q^2$--dependence of the solution are 
fully controlled by the ``genuine'' 
BFKL saddle-point at $\lambda_0=1/2$.

\begin{figure}
\begin{center}
\includegraphics[width=.75\textwidth] {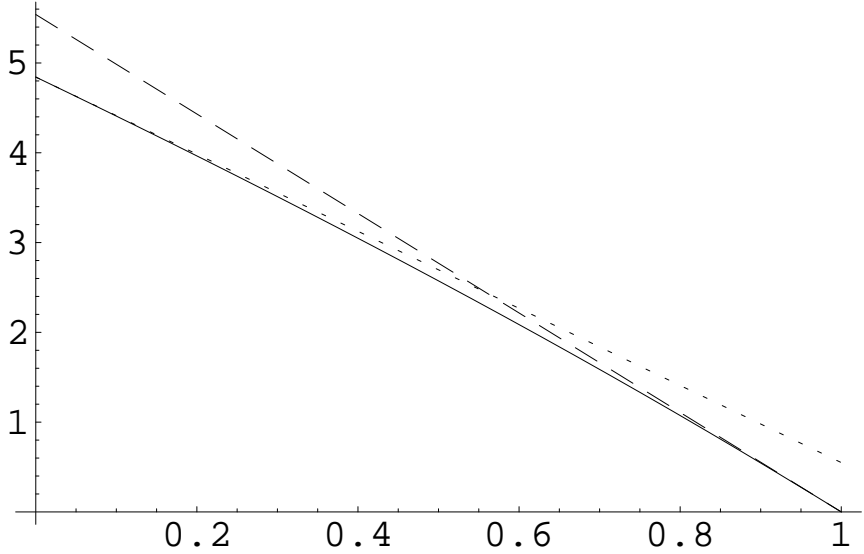}
\caption{The function $\rho_s(\tau,b)$, eq.~(\ref{rho_s_exact}),
together with small--$b$ approximation, eq.~(\ref{rhosmall})
(dotted line), and its large--$b$ approximation,
cf. eq.~(\ref{Qs_asy}) (dashed line) plotted as functions of $b/R_H(\tau)$.}
\label{2exp}
\end{center}
\end{figure}

The change of behaviour from eq.~(\ref{Qscenter}) to eq.~(\ref{Qs_asy})
is also visible on a logarithmic plot of the saturation scale in
eqs.~(\ref{rho_s_exact})--(\ref{Qs_exact}) as a function
of $b$. In Fig. \ref{2exp}, we have displayed
the function $\rho_s(\tau,b)$ of
eq.~(\ref{rho_s_exact}) as a function of $b/R_H(\tau)$, together
with its small-distance and long-distance approximations, as
given by eq.~(\ref{rhosmall}), and prior to eq.~(\ref{Qs_asy}),
respectively. As explicit on this figure, the transition between
the two regimes is rather smooth, and takes place at intermediate values
$b\sim R_H(\tau)/2$.

\subsection{Expansion of the black disk}

Let us finally consider the implications of the previous results on 
$Q_s$ for the expansion of the black disk. This is interesting since,
as we shall discover in Sect.~5 below, the two domains (I) and (II)
are characterized by different ``geometric scaling''
laws for  the black disk radius 
$ R(\tau,Q^2)$, and thus for the dipole cross-section.

Equation~(\ref{Qscenter}) 
together with the definition (\ref{condR}) of $R(\tau,Q^2)$ imply
(recall that $\gamma= 1/\lambda_s$):
\be\label{BDI}
2m_\pi R(\tau,Q^2)\Big|_{(I)}\,\simeq\,\lambda_s\bigg(c\bar\alpha_s \tau\,-\,
\ln \frac{Q^2}{\Lambda^2}\bigg)\,.\ee
This is valid as long as $R(\tau,Q^2)$ satisfies the condition
(\ref{smallb}), in practice, for
$R(\tau,Q^2)\simle  R_H(\tau)/2$. For a given $Q^2\gg \Lambda^2$,
this happens only within an intermediate
range of energies, to be specified shortly. Of course,
eq.~(\ref{BDI}) is just an approximate form of the general
expression (\ref{BDR_full}) valid in this intermediate
range of energies, but the approximations necessary to derive
eq.~(\ref{BDI}) may not be easily recognized at the
level of eq.~(\ref{BDR_full}). These approximations 
will be clarified in Sect. 5 below.

But at sufficiently high energy, the border of the black disk lies
in domain (II), that is, it is relatively close to the edge of the
hadron, in the sense of eq.~(\ref{largeb}). To verify this,
one can use eq.~(\ref{RR}) to deduce that
the ratio:
\be\label{RRratio}
\frac{R_H(\tau)-R(\tau,Q^2)}{R_H(\tau)}\,\approx\,
\frac{1}{2\omega\bar\alpha_s\tau}\,\ln \frac{Q^2}{\Lambda^2}\,,
\ee
is decreasing with $\tau$, and therefore necessarily satisfies
the condition (\ref{largeb}) for sufficiently large $\tau$.
Also, it can be easily checked that the expressions (\ref{Qs_asy}) for the 
saturation scale in domain (II) and (\ref{BD_asy}) for 
the black disk radius at high energy are consistent with each other,
via eq.~(\ref{condR}). This is as it should since both expressions have
been obtained via the same high energy approximation, namely, they follow 
from the asymptotic form (\ref{BFKL_sol_asy}) of the BFKL solution.

\begin{figure}
\begin{center}
\includegraphics[width=.9\textwidth] {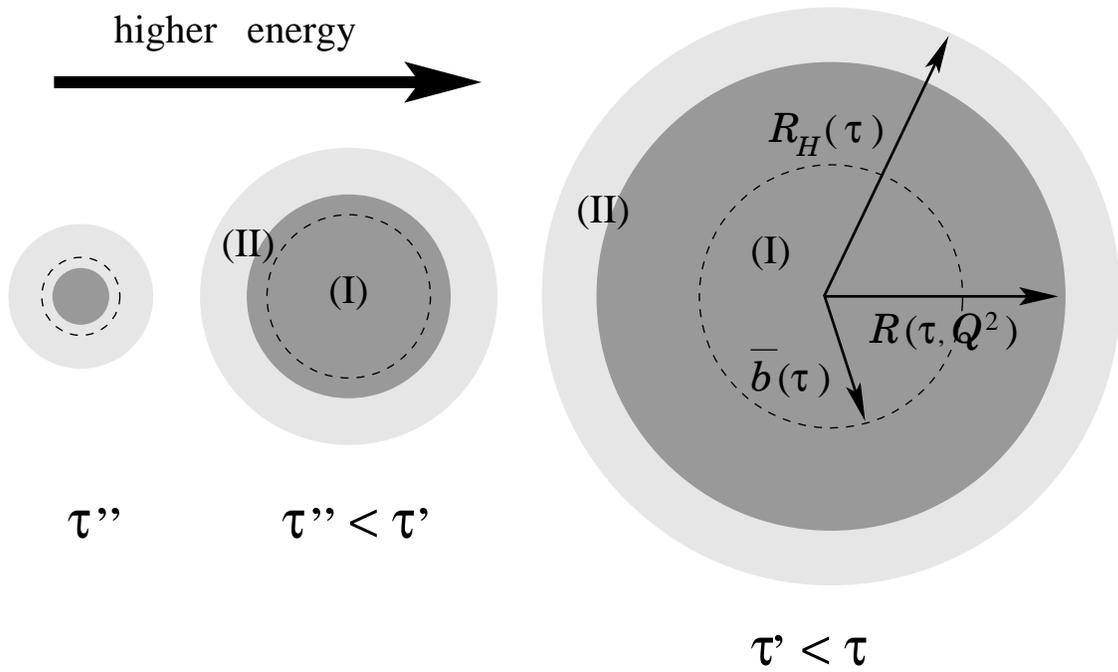}
\caption{A pictorial representation of the expansion of the black disk
with increasing rapidity. The dotted line circle of radius 
$\bar b(\tau)=R_H(\tau)/2$ separates between domains (I) and (II).}
\label{2BD}
\end{center}
\end{figure}

The evolution of the black disk with increasing $\tau$ at fixed $Q^2$
is pictorially illustrated in Fig. \ref{2BD}. The black disk appears
at the center of the hadron at a critical rapidity $\bar\tau_1(Q^2)$
such that $Q_s^2(\bar\tau_1,b=0) =Q^2$. This condition
together with eq.~(\ref{Qs0}) implies:
\be\label{criticial1}
\bar\tau_1(Q^2)\,=\,
\frac{1}{c\bar\alpha_s}\,
\ln \frac{Q^2}{\Lambda^2}\,.\ee
 For $\tau>\bar\tau_1(Q^2)$, but
not much larger, the black disk remains confined to domain (I), as
illustrated by the smallest disk on the
left of Fig. \ref{2BD}. 
However, the expansion rate of the black disk, which is
equal to $c\lambda_s\approx 3.12$ in appropiate units
(cf. eq.~(\ref{BDI})), is faster than the corresponding rate
$\omega/2\approx 1.39$ for the borderline $\bar b(\tau) \equiv
R_H(\tau)/2$ between the two domains. Thus, at
some new, larger, critical value, which can be easily estimated from
eq.~(\ref{BDI}) as
\be\label{criticial2}
\bar\tau_2(Q^2)\,=\,
\frac{1}{(c-\omega/2\lambda_s)\bar\alpha_s}\,
\ln \frac{Q^2}{\Lambda^2}\,,\ee
the black disk reaches domain (II), and than extends further within
this domain (as illustrated by the two larger disks in
Fig. \ref{2BD}). But within domain (II), the expansion rate
of the black disk slows down to $\omega\approx 2.77$,
cf. eq.~(\ref{BD_asy}),
which is the same rate as for the hadron outer border $R_H(\tau)$.
Thus, with increasing $\tau$, the radial distance between the border
of the black disk and the edge of the hadron (i.e., the width of
the grey area) remains constant (cf. eq.~(\ref{RR})), so the relative
size of the grey area with respect
to the hadron size is smaller and smaller (cf. eq.~(\ref{RRratio})).
At large $\tau$, this grey area represents the
 {\it tail} of the hadron wavefunction, in which the BFKL
equation remains valid even at arbitrarily large energy.

Given the (quasi)exponential decrease of the saturation scale
$Q_s^2(\tau,b)$ with $b$, it is straightforward to verify
that  the conditions $\Lambda^2\ll Q_s^2(\tau,b)\ll Q^2$ hold
at any point $\bb$ in the corona 
$R(\tau,Q^2) \ll \bb \ll R_H(\tau)$, as necessary for the consistency
of our approximations. Indeed, these conditions are satisfied
as soon as the separation
between $\bb$ and the (inner or outer) edges of this corona is of 
order $1/m_\pi$ or larger.
For instance, eq.~(\ref{Qs_asy}) can be rewritten
as:
\be
Q_s^2(\tau,b)\,\simeq \,\Lambda^2\, {\rm e}^{4m_\pi(R_H(\tau)-b)}\,.\ee
Similarly, by using eq.~(\ref{condR}), the expressions
(\ref{Qscenter}) and (\ref{Qs_asy}) can be recast into the form:
\be
Q_s^2(\tau,b)\,\simeq \,Q^2\,{\rm e}^{-2\gamma m_\pi (b-R(\tau,Q^2))}\,,
\ee
where $\gamma\approx 1.55$ if $\bb$ is in domain (I), and $\gamma=2$
for $\bb$ in domain (II).

\section{Geometric scaling at high energy}
\setcounter{equation}{0}

``Geometric scaling'' refers to the property of the dipole-hadron
total cross section  $\sigma(\tau,Q^2)$, eq.~(\ref{sigma}), to
depend upon the two kinematical variables $\tau$ and
$Q^2$ only via the combination $Q^2_0(\tau)/Q^2$ (the ``scaling
variable'') where $Q^2_0(\tau)$ is some suitable momentum scale which
increases as a power of the energy: $Q^2_0(\tau)\propto 
{\rm e}^{\lambda\tau}$. This property 
is interesting since it can be related to
a similar property of the virtual photon
total cross section $\sigma_{\gamma^* p}(\x,Q^2)$ which is
actually seen in the
HERA data on deep inelastic scattering 
(for  $\x < 0.01$ and $Q^2<400~{\rm GeV}^2$) \cite{geometric}.

Clearly, this property cannot hold for arbitrary $\tau$ and $Q^2$,
since it is known to be violated by the solutions to 
the linear evolution equations (DGLAP or BFKL) at very high $Q^2$.
On the other hand, in the saturation regime at $Q^2 <Q_s^2(\tau)$,
this property is physically motivated, since
the saturation scale is then the only scale in the 
problem \cite{MV94} (at least for a homogeneous hadron), so it
is tempting to identify it with the momentum scale $Q^2_0(\tau)$ 
introduced above. 

So far, studies of geometric scaling have been performed only for a 
homogeneous hadron, so they have naturally focused on the corresponding
 property of the scattering amplitude ${\cal N}_\tau(Q^2)$
(because $\sigma(\tau,Q^2)=2\pi R^2 {\cal N}_\tau(Q^2)$ in this
case). Such previous studies --- which relied on either
numerical \cite{LT99,Motyka,L01},  or (approximate)
analytic \cite{SCALING,MuellerT02}
solutions to the BK equation --- not only  confirmed the 
existence of geometric scaling in the saturation regime, but also 
showed that this property extends up to momenta $Q^2$
considerably larger than $Q_s^2(\tau)$.
In particular, in Ref. \cite{SCALING},  the upper limit for
``extended scaling'' has been estimated as 
$Q^2_{max}\sim Q_s^4(\tau)/\Lambda^2$,
which is roughly consistent with the phenomenology \cite{geometric}
(see also below).

Our purpose in this section is to extend the analysis of 
Ref. \cite{SCALING} by taking into account the transverse
inhomogeneity in the hadron, within the formalism developed
in the previous sections of
this paper. Note that, since the impact parameter is
integrated over in the formula (\ref{sigma}) for  $\sigma(\tau,Q^2)$,
the connection between the scaling properties of the cross-section
and those of the  scattering amplitude ${\cal N}_\tau(\rr,\bb)$ 
becomes more subtle now. In particular, it is not a priori clear 
what should be the scale which plays the role of
$Q^2_0(\tau)$ in the scaling variable, or even whether geometric
scaling exists at all (since the  inhomogeneous problem
is not a single-scale problem any longer).

To progressively introduce the effects of the inhomogeneity, let us start
with a situation where the rapidity $\tau$ is small enough for
the hadron to look ``grey'' everywhere: $Q^2 > Q_s^2(\tau,b=0)$,
or $\tau < \bar\tau_1(Q^2)$, cf. eq.~(\ref{criticial1}).
In that case, and within the present approximations,
the scattering amplitude ${\cal N}_\tau(Q^2,\bb)$ is 
factorized at any $\bb$ :
${\cal N}_\tau(Q^2,\bb)\simeq{\cal N}_\tau(Q^2)S(\bb)$.
Thus, the
study of the scaling properties of the total cross-section:
\be\label{sigmaALLGREY}
\sigma(\tau,Q^2)\,\simeq\,2{\cal N}_\tau(Q^2)\int d^2\bb\,S(\bb)
\qquad ({\rm all\,\,grey}),\ee
reduces to the corresponding study of ${\cal N}_\tau(Q^2)$
in the homogeneous case \cite{SCALING}.
It is instructive to briefly rederive here the
main result in Ref. \cite{SCALING} (in a very schematic way): 

The  homogeneous  scattering amplitude  ${\cal N}_\tau(Q^2)$ is given 
by (\ref{BFKL_sol1}) with $b=0$.
As such, this formula shows no scaling. To see the scaling emerging,
we shall replace the arbitrary reference scale $\Lambda^2$ in this
equation with the saturation scale $Q_s^2(\tau)=\Lambda^2 
{\rm e}^{c\bar\alpha_s \tau}$ (which is the same as the central
saturation  scale $Q_s^2(\tau,b=0)$ in the inhomogeneous case; cf.
eq.~(\ref{Qs0})). We write:
\be\label{LQS} \ln \frac{Q^2}{\Lambda^2}\,=\,\ln \frac{Q^2}{Q_s^2(\tau)}
\,+\,\ln\frac{Q_s^2(\tau)}{\Lambda^2}\,=\,\ln \frac{Q^2}{Q_s^2(\tau)}
+\,c\bar\alpha_s \tau\,.\ee
Since, by construction, the saturation scale is such that the exponent
in  (\ref{BFKL_sol1}) (with $b=0$)  vanishes for ${Q^2}={Q_s^2(\tau)}$,
it is clear that, after the replacement (\ref{LQS}), we are left
only with 
terms involving, at least, one power of $\ln ({Q^2}/{Q_s^2(\tau)})$ :
\BQ\label{BFKL_QS}
{\cal N}_\tau(Q^2)\,=\,
\exp\left\{- \lambda_s\ln \frac{Q^2}{Q_s^2(\tau)}
-\frac{1}{2\beta\bar\alpha_s\tau}
\left(\ln \frac{Q^2}{Q_s^2(\tau)}\right)^2
\right\}.
\EQ
Here, $\lambda_s\approx 0.64$ has been generated as
$\lambda_s=1/2+c/\beta$ (cf. eqs.~(\ref{c}) and (\ref{rhosmall})).
So far, (\ref{BFKL_QS}) is just a rewriting of eq.~(\ref{BFKL_sol1}).
But this is suggestive of the conditions under which
geometric scaling should be expected: This emerges when 
the (central) saturation scale $Q_s^2(\tau)$ is sufficiently close
to $Q^2$ (although still smaller than it) for the quadratic term
in eq.~(\ref{BFKL_QS}) to be negligible compared to
the linear term. When this happens, eq.~(\ref{BFKL_QS}) can be
approximated as:
\BQA
{\cal N}_\tau(Q^2)
&\simeq& \left(\frac{Q_s^2(\tau)}{Q^2}\right)^{\lambda_s}
\label{scaling_BFKL}
\EQA
which shows geometric scaling indeed, with
$Q_0^2(\tau)\equiv Q_s^2(\tau)$. This is valid as long as:
\be \label{window1}
1\,<\,\ln \frac{Q^2}{Q_s^2(\tau)}\,\ll\,2\lambda_s\beta\bar\alpha_s\tau,\ee
where the lower limit is simply the condition that we are in a
regime where eq.~(\ref{sigmaALLGREY}) applies: the hadron looks ``grey''
everywhere. Since $2\lambda_s\beta\approx 43.37$ is a 
large number, eq.~(\ref{window1}) gives a  rather
large window, which however extends beyond the validity range 
of the BFKL saddle point (\ref{BFKL_SP}) \cite{SCALING}. A more complete
analysis \cite{SCALING}  shows that eq.~(\ref{window1}) should
be replaced by\footnote{Note that the strong inequality on the logarithm 
in eq.~(\ref{window1}) has been replaced in eq.~(\ref{window2})
by a normal inequality,
which is equivalent to a strong inequality on the argument of the log.
This is the condition
$Q^2\ll Q_s^4(\tau)/\Lambda^2$ alluded to before \cite{SCALING}.}:
\be \label{window2}
1\,<\,\ln \frac{Q^2}{Q_s^2(\tau)}\,<\,\ln\frac{Q_s^2(\tau)}{\Lambda^2}\,
=\,c\bar\alpha_s \tau\,,\ee
which for the present purposes is rewritten as a range for $\tau$:
\be\label{Sgrey}
\frac{1}{2c\bar\alpha_s} \ln \frac{Q^2}{\Lambda^2}
\,<\,\tau\,<\,\frac{1}{c\bar\alpha_s} \ln \frac{Q^2}{\Lambda^2}\,.\ee
The lower limit in the equation above, which arises from the
upper limit in eq.~(\ref{window2}), is the smallest value of $\tau$
at which the BFKL solution starts to behave like a scaling function,
for a given $Q^2$. As for the upper limit --- in which we recognize 
the critical value $\bar\tau_1(Q^2)$ for the
emergence of the black disk, eq.~(\ref{criticial1}) ---, this is
necessary only for the validity of eq.~(\ref{sigmaALLGREY}).
For higher values of  $\tau$, 
geometric scaling may still hold, but in order
to see it, the calculation
should be modified to account for the formation of the black disk.
  
Specifically, for $\tau > \bar\tau_1(Q^2)$, the
cross-section (\ref{sigma}) can be then decomposed into a
``black'' contribution plus a ``grey'' one, which are evaluated as
(for $R(\tau,Q^2) > 1/m_{\pi}$):
\be\label{b+g}
\sigma(\tau,Q^2)&\simeq&2\pi R^2(\tau,Q^2)\,+\,
2{\cal N}_\tau(Q^2)\int d^2\bb\,S(\bb)\,\Theta(b-R(\tau, Q^2))\nn
&\simeq&2\pi R^2(\tau,Q^2)\,+\,
\frac{2\pi R(\tau, Q^2)}{m_{\pi}}\,.\ee
The second line is obtained from the first one after replacing  
$S(b)\approx {\rm e}^{-2m_{\pi}b}$, and noticing that 
${\cal N}_\tau(Q^2){\rm e}^{-2m_{\pi}R(\tau,Q^2)}=1$.
Eq.~(\ref{b+g}) confirms that, when the energy is large enough
for $R(\tau,Q^2) > 1/m_{\pi}$, the total cross-section is 
dominated by the black disk, as anticipated in Sect. 2.
It further shows that the
scaling properties of the cross section are determined by the 
corresponding properties of the radius of
the black disk, which can be inferred from the discussion
in Sect. 4. 
As in Sects. 4.3--4.4, we are led to distinguish between two regimes:
\bigskip

I) If $\bar\tau_1(Q^2)<\tau <\bar\tau_2(Q^2)$, 
cf. eqs.~(\ref{criticial1})--(\ref{criticial2}),
the black disk lies entirely inside domain (I), and
the corresponding radius is given by eq.~(\ref{BDI}), which
is now rewritten as\footnote{Incidentally, 
this also shows that the way to derive eq.~(\ref{BDI}) from 
the general expression (\ref{BDR_full}) is via manipulations similar
to those leading from eq.~(\ref{BFKL_sol1}) to
eq.~(\ref{scaling_BFKL}), cf. eqs.~(\ref{LQS})--(\ref{BFKL_QS}). } :
\be\label{BDI1}
R(\tau,Q^2)\Big|_{(I)}\,\simeq\,\frac{\lambda_s}{2m_\pi}
\ln \frac{Q_s^2(\tau,b=0)}{Q^2}\,.\ee
This shows scaling,
with the same scaling variable $Q^2_s(\tau)/Q^2$ as in the
``all grey'' regime (\ref{Sgrey}). 

II) In the high-energy regime $\tau > \bar\tau_2(Q^2)$,
the edge of the black disk
is in  domain (II), so its radius is given by eq.~(\ref{BD_asy}).
This shows scaling too, but with a different
scaling variable:
\BQ
R(\tau,Q^2)\Big|_{(II)}
\,=\,\frac{1}{4m_\pi} \ln \frac{ Q^2_{\infty}(\tau)}{Q^2},\qquad
 Q^2_{\infty}(\tau)\equiv 
\Lambda^2 {\rm e}^{2\omega \bar\alpha_s\tau}. 
\EQ
This gives the scaling law for the total cross section at 
very high energy:
\BQ\label{Fscaling}
\sigma(\tau,Q^2)\simeq 2\pi R^2(\tau,Q^2)= \frac{\pi}{2m_\pi^2}
\left(\ln \frac{Q^2_{\infty}(\tau)}{Q^2}\right)^2 \, .
\EQ 

To summarize, 
for fixed $Q^2$ and intermediate energies corresponding to the
following  range of rapidities (cf. eqs.~(\ref{Sgrey}) and
(\ref{criticial2})):
\be\label{SI}
\frac{1}{2c\bar\alpha_s} \ln \frac{Q^2}{\Lambda^2}
\,<\,\tau\,<\,\frac{1}{(c-\omega/2\lambda_s)\bar\alpha_s}\,
\ln \frac{Q^2}{\Lambda^2}\,=\,\bar\tau_2(Q^2)\,,\ee
the dipole-hadron cross-section exhibits geometric scaling,
with the scale set by the {\it central} saturation scale
$Q^2_s(\tau)\equiv Q^2_s(\tau,b=0)$. Eq.~(\ref{SI})
allows for a significant window since, numerically, 
$2c\approx 9.68$ while $c-\omega/2\lambda_s \approx 2.69$.
On the other hand, at higher energies, corresponding to 
 $\tau > \bar\tau_2(Q^2)$, there is scaling again,
but the relevant scale is rather the {\it asymptotic} scale
$Q^2_{\infty}(\tau)$.

Of course, the scaling breaks down, strictly speaking,
for $\tau \sim \bar\tau_2(Q^2)$, i.e., in  the transition regime
where the black disk crosses from domain (I) to domain (II). But it
so happens that, numerically, the difference between
the exponents $c\approx 4.84$ and
$2\omega\approx 5.55$ in the  corresponding scaling variables
($Q^2_s(\tau)$ and, respectively, $Q^2_{\infty}(\tau)$) is quite
small, so the transition from one scaling law to the other
takes places rather fast, as also
manifest in the plot in Fig. \ref{2exp}.

It should be also stressed that, with incresing $\tau$, the
geometric scaling becomes less and less relevant, since the
$Q^2$--dependence of the cross-section (\ref{Fscaling})
eventually becomes a subleading effect: When the condition
(\ref{r/t}) is fulfilled, the leading-order term in eq.~(\ref{Fscaling})
is simply proportional to $\tau^2$, with the scale set by the pion
mass (cf. eq.~(\ref{sigmaFC})).

For completness, let us conclude with a discussion of the
scaling properties of the {\it local} scattering amplitude 
${\cal N}_\tau(Q^2, \bb)$. Within the grey area,  the
factorization property 
${\cal N}_\tau(Q^2, \bb)=S(\bb){\cal N}_\tau(Q^2)$ implies
that any scaling property of the homogeneous solution 
${\cal N}_\tau(Q^2)$ transmits automatically to ${\cal N}_\tau(Q^2, \bb)$,
but with a scaling variable involving the {\it local} saturation scale
$Q_s^2(\tau,b)$ :
\BQA\label{Nbscaling}
{\cal N}_\tau(Q^2)
\,=\, \left(\frac{Q_0^2(\tau)}{Q^2}\right)^{\lambda}
\,\longrightarrow \,\,\,{\cal N}_\tau(Q^2, \bb)
\,=\, \left(\frac{Q_s^2(\tau,b)}{Q^2}\right)^{\lambda}\,,
\EQA
with $Q_s^2(\tau,b)=Q_0^2(\tau){\rm e}^{-2m_\pi b/\lambda}$.
This is precisely the scaling solution 
(\ref{Nscaling}) that we used in our arguments in Sect.~\ref{scat_grey}.
In eq.~(\ref{Nbscaling}), the momentum scale
$Q_0^2(\tau)$ and the power $\lambda$ depend upon the rapidity
$\tau$, in the expected way: When $\tau$ is 
in the intermediate range (\ref{SI}), $Q_0^2(\tau)$ is
the saturation scale
at the center $Q_s^2(\tau,b=0)$, and $\lambda=
\lambda_s\approx 0.64$. At higher rapidities, $\tau > \bar\tau_2(Q^2)$,
one rather has $Q_0^2(\tau)=Q^2_{\infty}(\tau)$ and $\lambda=1/2$.
  
In both cases,  eq.~(\ref{Nbscaling}) holds only in the grey
area at $b > R(\tau, Q^2)$. But inside the black disk, geometric
scaling holds as well, and almost trivially, since there
${\cal N}_\tau(Q^2, \bb)\approx 1$ and the deviation from
one shows scaling too \cite{LT99,SAT,SCALING}. We thus conclude
that, for any $\tau$ above the lower limit in eq.~(\ref{Sgrey}),
the scattering amplitude ${\cal N}_\tau(Q^2, \bb)$ shows
geometric scaling everywhere in the hadron disk.

\section{Conclusions and outlook}

{ In this paper, we have proposed a simple mechanism combining perturbative
gluon saturation and non-perturbative boundary conditions which ensures
the saturation of the \FB in dipole-hadron scattering at high energy.
Gluon saturation has been implemented via the non-linear
evolution equations derived in perturbation theory
in Refs. \cite{B,K,W,PI}, that we have used only for impact parameters
in the central region of the hadron, where the gluon density is high
and perturbation theory is applicable. 
On the basis of these equations, we have shown that, 
with increasing energy, the non-linear effects
ensure not only the unitarization of the scattering amplitude at
fixed impact parameter, 
but also the factorization of the impact
parameter dependence of the scattering amplitude in the outer 
``grey area'', where the unitarity limit has not yet been reached. 
This factorization, together with the exponential
fall-off of the non-perturbative initial condition at large distances,
leads to a total cross-section which grows like $\ln^2 s$.
The coefficient of this growth is universal, i.e., independent
of the hadronic target, and has been computed here.}

Our analysis makes explicit the deep connection between
unitarization effects and the formation of the Colour Glass Condensate.
The ``black disk'' within which the unitarity limit has been reached
is precisely the region of the hadron where the gluons ``seen'' by the
incoming dipole are saturated, whereas the ``grey area'' corresponds
to a region of lower density, in which the BFKL evolution still applies.
The transition from ``grey'' to ``black'', i.e., the expansion
of the black disk, is described by the non-linear evolution equations
for the Colour Glass Condensate. The colour correlations
at saturation are essential for both factorization and 
unitarization: They ensure colour
neutrality for the saturated gluons, and thus suppress the non-unitary
contributions due to the long-range Coulomb tails. 

We have given the first computation of the saturation scale for
an inhomogeneous hadron and identified two factorization
regimes at  different impact parameters. This has interesting
consequences for the geometric scaling properties of the
total cross-section: Our analysis predicts two different scaling
laws in different ranges of energy. 

For simplicity, our analysis has been carried out for an external
probe which is a $q\bar q$ dipole of fixed transverse size. 
But its extension to deep inelastic scattering should be
straightforward. To this aim, the 
dipole-hadron cross-section must be averaged over all
transverse sizes of the $q\bar q$ pair, with a weight given by the
light-cone wavefunction of the virtual photon \cite{AM0,AM3,NZ91}.

It is important to point out two limitations of our approach,
which call for further studies. First, it was essential for our
approximation scheme that the dipole is small, $\rr\ll 1/\Lambda_{QCD}$.
In particular, this restricts the applicability of our results
to deep inelastic scattering at relatively large photon virtuality,
$Q^2 \gg \Lambda_{QCD}^2$. 
It would be most interesting to see if one can
phenomenologically  extend this analysis towards the strong 
coupling regime at $Q^2 \sim \Lambda_{QCD}^2$ using
the "soft pomeron", and thus get at least an
estimate for the coefficient of the $\ln^2 s$ growth in that regime.

Second, we have performed our analysis at fixed coupling. It is an
interesting open question 
how our results would be modified by  the running
of the coupling. Strictly speaking, this running
is one of the next-to-leading order effects which
so far have not been systematically included in the non-linear
evolution equations (see, however, \cite{BB01}). But the experience
with the BFKL equation, for which the NLO corrections have been
recently computed \cite{NLBFKL}, suggests that 
a significant part of these corrections could be indeed taken
into account by including the running of the coupling. 
Then, the natural question is, what should be the scale at which
the coupling must be evaluated. 

In a previous paper \cite{SCALING}, where the saturation scale
$Q_s(\tau)$ for a {\it homogeneous} hadron has been computed,
we found it natural to evaluate the coupling at the 
saturation momentum. Indeed, in that problem, the rapidity
$\tau$ and the external momentum $Q^2$ (the resolution of the
dipole) were increased {\it simultaneously}, 
in such a way to preserve
the saturation condition $Q_s^2(\tau)=Q^2$. For that particular
running, we have shown that
the saturation momentum changes its parametric dependence upon
rapidity with respect to the fixed coupling case,
from ${\rm e}^{c\alpha_s \tau}$ to
${\rm e}^{\kappa \sqrt{\tau + \tau_0}}$ (with 
constant $\kappa$ and $\tau_0$). This result has been
subsequently  confirmed in Ref. \cite{MuellerT02}.

On the other hand, in the present paper the physical situation
is quite different: The  external momentum $Q^2$ is {\it fixed}, 
and with increasig $\tau$, we simultaneously increase the impact
parameter $b$, in such a way that the condition $Q_s^2(\tau,b)=Q^2$
(which defines the edge of the black disk) remains satisfied.
It seems therefore natural to evaluate the coupling at the 
fixed external scale $Q^2$, in which case the fixed coupling expressions 
obtained in this paper would remain valid (after the trivial replacement
$\alpha_s\to \alpha_s(Q^2)$). A further argument in this sense
is provided by the discussion in Sect. 3.1, which shows that the
dominant scattering involves only nearby colour sources, within the
area covered by the incoming dipole or slightly further away,
so that the typical transferred  momenta are of order  $Q^2$.

A still different possibility for the running, 
which would be closer in spirit to Ref. \cite{SCALING}, would be to 
evaluate the coupling at the {\it local} saturation scale $Q_s^2(\tau,b)$.
Although this scale and the external scale $Q^2$ are identified at the
edge of the black disk, they are nevertheless different at the
points $\bb$ which lie further away in the grey area 
(where $Q^2\gg Q_s^2(\tau,b)$). Thus, this choice for the running 
would probably modify the current formulae, in such a way to provide a
generalization of the results in Ref. \cite{SCALING} to the case of
an inhomogeneous hadron.

Of course, the description above is extremely crude and exploratory,
and a full analysis of the running
coupling case is required before any strong conclusions can be drawn.

\vspace*{.5cm}
\section*{Acknowledgments}

Kazunori Itakura and Edmond Iancu would like to thank the hospitality of
KITP at the University of California in Santa Barbara, where this work was 
completed, and the organizers of the program {\it ``QCD and Gauge 
Theory Dynamics in the RHIC Era''}. We also thank Dima Kharzeev
for useful discussions, and Genya Levin and Basarab Nicolescu for helpful
comments on the original version of the manuscript.
Larry McLerran would like to acknowledge useful 
conversations with Henri Kowalski, Al Mueller and Raju
Venugopalan. This manuscript has been authorized under Contracts No.
DE-AC02-98CH10886 and No. DE-AC02-76CH0300 with the U.S. Department of
Energy. 
This research was supported in part by the National Science Foundation 
under Grant No. PHY99-07949. 

\vspace*{.5cm}
\section*{\centerline{Note added}}

After the submission of the original version of this paper, there appeared a
preprint by Kovner and Wiedemann which disputes the claims of this paper
\cite{KW022}. In Sec. 3.2, we have added some clarifying comments which 
we hope will ameliorate this dispute. In this note, we would like to discuss
this in more detail.

We are first of all pleased to notice that the authors of 
Ref. \cite{KW022} have to some extent agreed with our criticism 
to their papers \cite{KW02}, and modified their arguments accordingly.
For instance, they now agree that the dominant contribution to the 
scattering within the ``grey area'' comes from local sources (although
their exact definition of the ``grey area'' is somewhat different from ours).
Also,  they admit that the long-range,
non-unitary, contribution to the scattering amplitude that they previously
computed in  Ref. \cite{KW02} applies only for a {\it coloured} probe, and
not for a dipole.

But in spite of their agreement on these points, the authors of 
Ref. \cite{KW022} dispute our conclusion on ``Froissart bound from gluon 
saturation'', and instead conclude that the Froissart bound is 
violated by the perturbative evolution equations in Refs. \cite{B,K,W,PI}.
In its essence (but using the language of our Sect. 3.2), 
the argument of Ref. \cite{KW022} can be formulated as follows: 
Assume that one starts the perturbative evolution
at some early rapidity $\tau_0$, at which 
the point $\bb$ of interest lies far outside the initial ``grey area'' 
($b \gg R_H(\tau_0)$), i.e., in the ``white area'', according to the
terminology of Ref. \cite{KW022}. In this perturbative setting, the
dominant contribution to the scattering amplitude at $\bb$
is the {\it long-range} contribution, which decreases only as 
$1/\bb^4$, cf. eq.~(\ref{Ngrey}). With increasing $\tau$, the central
region expands, and the point $\bb$ is eventually incorporated in the
``grey area'' at some ``time'' $\tau_1$. For $\tau>\tau_1$, the
evolution proceeds locally at $\bb$, but with an initial condition
at $\tau=\tau_1$ which decreases as a power of $\bb$, rather than as
an exponential. Thus, the first term in the r.h.s. of eq.~(\ref{Ngrey})
is replaced by $
{\rm e}^{\omega\bar\alpha_s\tau}/\bb^4$, and the ``blackness'' condition
${\cal N}_\tau(\rr,\bb)\sim 1$ yields a black disk increasing exponentially
with $\tau$. 

Although mathematically correct (at least, in its above formulation;
we disagree with some of the intermediate steps and detail points in
Ref. \cite{KW022}), this argument is physically irrelevant: It simply
signals a pathology of perturbation theory when this is used in a region
where it is not supposed to work, namely, in the ``white area'', where
physics is truly controlled by the confinement.
In the discussion in Sect. 3.2, we pointed out that, under plausible
assumptions about the non-perturbative physics in this region, 
the true evolution respects the exponential fall-off of the initial condition.

In any case, our point in this paper
was {\it not} to argue that, by integrating the perturbative evolution
up to {\it arbitrarily} large $\tau$, one would generate cross-sections
which saturate the \FB indefinitely. Rather, we have shown that by applying
perturbation theory where it is expected to work --- namely,
in the central region at $b < R_H(\tau)$, where the condition
$Q_s(\tau,b) \gg \Lambda_{QCD}$ is satisfied by definition
(cf. eq.~(\ref{RH})) ---, one finds that
({\it i}\,) it is possible to compute the expansion rate of the black disk,
and ({\it ii}\,) this rate is such that the \FB is saturated. 
None of these points is trivial:

({\it i}\,) It was not a priori obvious that the expansion of the
black disk is controlled by perturbation theory. We have demonstrated this
for the case of a {\it small} dipole ($Q^2\gg \Lambda_{QCD}^2$), for which
we have shown that the ``grey area'' ($\equiv $ the region in which
$Q_s^2(\tau,b)$ decreases from $Q^2$ down to $\Lambda_{QCD}^2$) has a
considerable extent (cf. eq.~(\ref{RR})). This is the region which controls
the expansion, and perturbation theory is applicable here.

({\it ii}\,) It was  not a priori clear that the perturbative evolution of
the black disk will respect the \FBB. Indeed, the saturated gluons from the
black disk generate long-range forces which remain perturbative at impact
parameters within the ``grey area'', and thus cannot be discarded by invoking
confinement. (This was precisely the objection to \FB raised
in Ref. \cite{KW02}.)
We have shown that these long-range forces, although present, 
are nevertheless harmless: Because the saturated gluons are globally 
colour-neutral (an aspect which has been overlooked in Ref. \cite{KW02}), 
they generate only {\it dipolar} forces, whose contribution
to the scattering amplitude in the grey area remains smaller than the 
corresponding contribution of the local sources (which
saturates the \FBB).

The only place where non-perturbative physics has entered our argument
was in providing the $\bb$--dependence of the initial condition for the
short-range contribution. On physical grounds, this is fixed by confinement,
and thus is certainly of the exponential type (and not of the power-law type,
as a na\"{\i}ve perturbative evolution starting in the ``white area''
would predict).

To conclude, Kovner and Wiedemann seem to be addressing a
different question:  Whether or not one can na\"{\i}vely apply the
perturbative evolution equation in all regions of impact parameter space ---
including over distances many times the pion Compton wavelength ---, 
and still get sensible results.  They negatively answer this question,
and we definitely agree with them on this point.
On the other hand, they further make
the statement that one cannot compute the high-energy cross section using the
ideas associated with (perturbative) saturation.  
The body of this paper clearly contradicts this claim.


\end{document}